\documentclass[12pt,english]{article}
\usepackage[T1]{fontenc}
\usepackage[latin9]{inputenc}
\usepackage{geometry}
\usepackage{color}
\usepackage{amsmath}
\usepackage{amssymb}
\usepackage{graphicx}
\usepackage{setspace}
\makeatletter

\providecommand{\tabularnewline}{\\}

\@ifundefined{date}{}{\date{}}
\makeatother

\usepackage{babel}
\begin{document}

\title{Monomer Fluctuations and the Distribution of Residual Bond Orientations
in Polymer Networks}

\author{Michael Lang}
\maketitle
\begin{abstract}
In the present work, four series of simulations are analyzed: entangled
model networks of a) mono-disperse or b) poly-disperse weight distribution
between the crosslinks, c) non-entangled phantom model networks and
d) non-entangled model networks with excluded volume interactions.
Previous work on the average residual bond orientations (RBO) of model
networks \cite{key-22} is extended to describe the distribution of
RBOs for the entangled networks of the present study. The phantom
model can be used to describe the monomer fluctuations, average RBOs,
and the distribution of RBOs in networks without entanglements and
without excluded volume. Monomer fluctuations in networks with excluded
volume but without entanglements can be described, if the phantom
model is corrected by the effect of the incompletely screened excluded
volume. It is shown that parameters of the tube model can be determined
from monomer fluctuations in polymer networks and from the RBO. A
scaling of the RBO $\propto N^{-1/2}$ is observed in both mono-disperse
and poly-disperse entangled networks, while for all networks without
entanglements a RBO $\propto N^{-1}$ is found. The distribution of
the RBO of entangled samples can be described by assuming a normal
distribution of tube lengths that is broadened by local fluctuations
in tube curvature. Both observables, monomer fluctuations and RBO,
are in agreement with slip link or slip tube models \cite{key-14}
for networks and disagree with network models that do not allow a
sliding motion of the monomers along a confining tube.
\end{abstract}

\section{Introduction}

To understand the effect of entanglements between polymer chains is
one of the main open problems of polymer physics \cite{key-37}: entanglements
control the rheology of high molecular weight polymer liquids and
lead to different relaxation functions depending on the architecture
of the molecules. Entanglements contribute to the modulus of polymer
networks and thus, affect swelling or deformation of networks and
gels \cite{key-11}. For modeling entanglements, concepts like a confining
tube were introduced \cite{key-40} to capture the collective effect
of all entanglements with the surrounding molecules \cite{key-24}.
But numerous experiments revealed that additional contributions are
necessary to model the effect of entanglements \cite{key-14,key-38},
for instance, rearrangements of the tube, fluctuations of the chains
inside the tube, or a change in tube diameter upon deformation.

The main difficulty is that the nature of the confinement is determined
indirectly from experimental data. In the case of computer simulations,
additional methods can be constructed that either require to modify
the sample to determine the ``primitive path'' of the tube, or use
a static construction method without considering dynamics (see \cite{key-39}
for a review of these methods), or superpositions of different trajectories
or conformations \cite{key-52}. Slip link models \cite{key-53,key-54}
or similar approaches aim to project this problem into binary interactions
among neighboring chains. It is not clear to date whether such a binary
or a more collective approach like a tube are more successful in all
relevant experimental situations.

The present work uses assumptions consonant with a slip tube model
\cite{key-14} for polymer networks to compute monomer fluctuations
and the residual bond orientations (RBO, \cite{key-73}). This latter
quantity is accessible by nuclear magnetic resonance (NMR) and therefore,
establishes a link between simulation and experiment. The RBOs reflect
both static and dynamic properties of the polymers \cite{key-8} and
therefore, require always a simultaneous analysis of conformations
and fluctuations to test any theoretical model. For polymer networks,
NMR has achieved large practical and theoretical importance, since
it can be used to characterize segmental order \cite{key-3,key-4}
and cross-link densities \cite{key-5,key-6,key-59} for a quick access
to basic elastic and dynamic properties of the networks.

Recently, the phantom and affine model predictions for the average
RBOs were extended to entangled networks \cite{key-22}. It was shown
that the time average monomer fluctuations in a network can be interpreted
assuming random walk statistics of a confining tube, and thus, might
allow to extract parameters related to the entanglement of the chains.
The basic idea of this recent work is that chain segments slide along
a confining tube and thus, sample tube sections of different orientations.
This leads to an analysis of NMR data similar to polymer melts based
upon the reptation picture \cite{key-7}, but for networks, the sliding
length along the tube reaches a long time limit $\propto bN^{-1/2}$
for inner chain monomers. In total, an average RBO at long times is
predicted $\propto N^{-1/2}$ in contrast to the $1/N_{e}$ contribution
of entanglements to network elasticity. As a result, one must conclude
that NMR does not measure stress inside entangled networks.

From experimental side, the entanglement length is defined by a strand
of $N_{e}$ monomers that contributes an energy of $kT$ per volume
of $N_{e}$ monomers to the plateau modulus. In theory and for analyzing
simulation data it is preferable to directly connect to the properties
of the confining tube, the primitive path, or to the statistics of
slip links, since this allows for a better test of theory. Obviously,
relations between experimentally determined modulus and these parameters
are model dependent \cite{key-14,key-56,key-55}. The main differences
are that the contribution to modulus per correlated section of the
chain may become smaller than $kT$ due to rearrangements of the tube
or the slip-links, or that the interactions are essentially pairwise
(slip-links) instead of collective (tube). In order to obtain a \emph{model}
\emph{independent} analysis for the nature of the confinement, we
introduce here two additional degrees of polymerization that are directly
accessible in the simulations of the present study: a) the correlation
degree of polymerization $N_{p}$ of the primitive path, (or the mean
distance between slip-links) and b) a degree of polymerization $N_{c}$
related to the stretching of the chains inside the tube. Following
the discussion in Refs \cite{key-14,key-56,key-55}, we expect $N_{p},N_{c}<N_{e}$,
whereby the relation between these parameters requires a model of
rubber elasticity and a detailed model for entanglements.

\begin{figure}
\begin{center}\includegraphics[width=0.36\paperwidth]{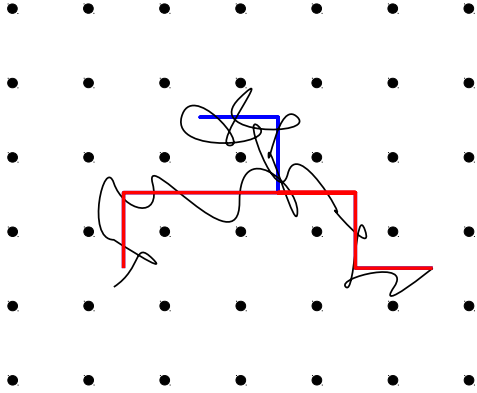}\end{center}

\caption{\label{fig:Sketch-of-an}Sketch of an entangled chain (black line)
in an array of obstacles. The red thick line shows the primitive path
of the chain, while the blue line shows entangled sections that are
not part of the primitive path.}

\end{figure}

The differences between $N_{c}$ and $N_{p}$ can be clarified using
Figure \ref{fig:Sketch-of-an}. Here, a polymer chain is shown in
an array of obstacles, whereby also the primitive path and all entangled
sections of the chain are shown. In our approach, $N_{c}$ describes
the stretching of the chain inside the tube, which is equivalent to
the number of entangled sections of the \emph{primitive path} ($N/N_{c}=6$
in Figure \ref{fig:Sketch-of-an}). $N_{p}$ on the other hand is
related to the average number of entangled sections \emph{per chain}
($N/N_{p}=10$ in Figure \ref{fig:Sketch-of-an}). For the model of
a single chain in an array of obstacles there must be $N_{p}\le N_{c}$
as shown by Helfand et al. \cite{key-27,key-74}, whereby a similar
behavior is expected \cite{key-74} for irregular arrays of obstacles
or entangled polymer melts. But without knowing the grid of obstacles,
one cannot distinguish subsequent parallel entangled sections. Thus,
determining $N_{p}$ from the correlation of entangled sections may
lead to a slight overestimate of the true $N_{p}$, while these correlations
do not shift $N_{c}$. Since the correlations are are limited by $N/N_{p}$,
we expect an exponential convergence from below towards the asymptotic
$N\rightarrow\infty$ limit, $N_{p,\infty}$, as function of $N/N_{p}$,
\begin{equation}
N_{p}(N)\approx N_{p,\infty}-c_{p}\exp(-N/N_{p}).\label{eq:Np}
\end{equation}
Here, $c_{p}$ is the numerical coefficient describing the amplitude
of this correction. For $N_{c}$, an additional stretching of the
chains is expected that arises for any stepwise cross-linking reaction
\cite{key-75,key104}. This stretching decays here for large $N$,
since the distortion of the conformations due to crosslinking is of
order $N^{1/2}$, while the length of the tube grows $\propto N$.
This leads to a convergence of $N_{c}$ as function of $\left(N/N_{c}\right)^{-1/2}$
from below towards the asymptotic limit $N_{c,\infty}$, 
\begin{equation}
N_{c}(N)\approx N_{c,\infty}-\frac{c_{c}}{\left(N/N_{c}\right)^{1/2}}.\label{Nc}
\end{equation}
Thus, $N_{e}$, $N_{p}$, or $N_{c}$ are not solely related by numerical
coefficients in polymer networks, if $N$ is finite. Note that a distinction
between $N_{p}$ and $N_{c}$ was not part of the preceding letter
\cite{key-22}. Thus, the analysis of the average RBO of Ref \cite{key-22}
is generalized in the present publication.

The structure of the paper follows the theoretical concepts and the
corresponding model networks. Therefore, most sections contain theory
\emph{and} simulation data. The reason for this choice is that any
later model is an improvement of the previous model and possible known
corrections - as shown by comparison with the simulation data - can
be introduced or neglected step by step.

Thus, the paper is structured as follows: the simulations are described
in section \ref{sec:Computer-simulations}. Monomer fluctuations and
the average RBO of non-entangled model networks is discussed in section
\ref{sec:Theory-and-Simulation}. The analogous discussion for entangled
model networks follows in section \ref{sub:Entangled-networks:-average}.
Then, theory is extended in section \ref{sub:Distribution-of-order}
to compute the distribution of RBOs in mono-disperse networks. The
model is extended to describe poly-disperse networks in section \ref{sec:Polydisperse-Networks}.
The fundamental differences to previous works are discussed in section
\ref{sec:Discussion}. More technical details related to time averaging,
the analysis of the networks, or finite size effects can be found
in the Appendix.

\section{Computer simulations\label{sec:Computer-simulations}}

\begin{figure}
\begin{center}\includegraphics[width=0.7\columnwidth]{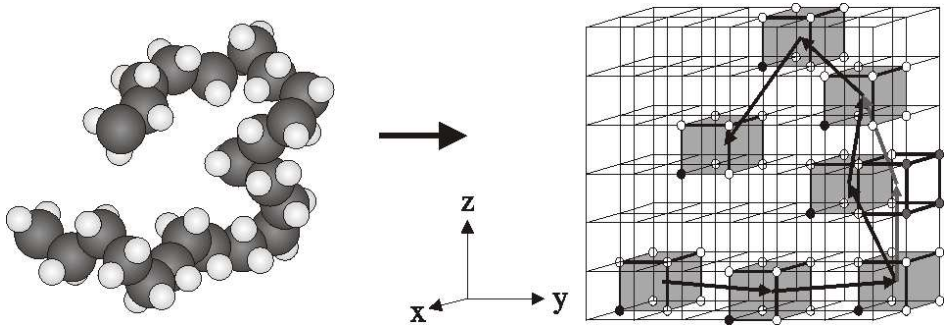}\end{center}

\caption{\label{fig:Bond-fluctuation-method.}Bond fluctuation method. Polymer
conformations are modeled by a chain of cubes on a cubic lattice and
dynamics is obtained by random hops of the cubes on the lattice as
indicated by the open cube on the right.}
\end{figure}

All simulations are conducted with a three dimensional version of
the bond fluctuation method (BFM) of Carmesin and Kremer \cite{key-15},
see Fig. \ref{fig:Bond-fluctuation-method.} for more details. Details
of network preparation can be found in Refs \cite{key-16,key-62}.
We focus on four series of simulations resulting in a) entangled mono-disperse
end-linked networks, b) entangled poly-disperse randomly cross-linked
networks, c) mono-disperse networks without entanglements and d) mono-disperse
networks with no excluded volume interactions and entanglements.

For the mono-disperse networks, mono-disperse melts of $M$ chains
of length $N=16,...,256$ (cf. Table \ref{tab:mono-1}) monomers were
equilibrated on a lattice of $128^{3}$ lattice sites at occupation
density $\phi=0.5$ in periodic boundaries. The equilibration time
was chosen such that at least 5 relaxation or reptation times per
chain \cite{key-23} could be realized. Afterward, $M/2$ of 4-functional
cross-links were inserted and relaxation of the chains continued.
A reaction took place whenever a free chain end diffused into the
nearest possible separation to a cross-link that was not yet fully
reacted. The reactions were stopped at 95\% degree of end-linking
in order to obtain samples with similar density of elastically active
material.

Two more series of simulations are derived from using identical copies
of the mono-disperse networks with $N=16$, 64, and 256 as initial
conformations. These copies were relaxed for $2\times10^{7}$ Monte-Carlo
steps (MCS) whereby excluded volume was maintained for one series
but the same extended bond vector set as in Ref. \cite{key-48} was
used to allow the strands to cross. For the other series, excluded
volume was switched off and thus, simultaneously all entanglements.
In order to provide an intuitive distinction between these different
network types, these networks are called ``phantom'' (no excluded
volume, no entanglements), ``crossing'' (no entanglements but excluded
volume), or ``entangled'' (entanglements and excluded volume) throughout
this work. Note that only the ``phantom'' networks correspond to
the phantom network model and that only the segments of these networks
are indeed Kuhn segments; the discussion of the ``crossing'' networks
in the following sections requires additional corrections \cite{key-42}
due to the incompletely screened excluded volume of the monomers.

\begin{table}
\begin{center}%
\begin{tabular}{|c|c|c|c|c|c|c|c|c|}
\hline 
$N$  & $M$  & $n$  & $bN_{p}^{1/4}$  & $N_{p}$  & $N_{c}$  & $m$  & $\phi_{\mbox{ac}}$  & $z$\tabularnewline
\hline 
\hline 
16  & 8192  & 2.32  & 4.45  & 8.2  & 6.5 & 0.068  & 0.878  & 0.413\tabularnewline
\hline 
32  & 4096  & 3.40  & 4.80  & 11.1  & 9.1 & 0.044  & 0.878  & 0.400\tabularnewline
\hline 
64  & 2048  & 4.47  & 4.94  & 12.4  & 11.4 & 0.032  & 0.882  & 0.389\tabularnewline
\hline 
128  & 1024  & 5.53  & 4.96  & 12.6  & 12.5 & 0.023  & 0.888  & 0.367\tabularnewline
\hline 
256  & 512  & 6.16  & 4.96  & 12.6  & 13.2 & 0.017  & $\approx0.91$  & 0.368\tabularnewline
\hline 
\end{tabular}\end{center}

\caption{\label{tab:mono-1}Networks contain $M$ chains of $N$ monomers.
$bN_{p}^{1/4}$ and virtual chain length $n$ are determined from
Figure \ref{fig:Monomer-fluctuations-in}. The length of a correlated
tube section $N_{p}$ is computed using $b=2.63$. The weight fraction
of active material $\phi_{\mbox{ac}}$ is used to extrapolate towards
a perfect network with no defects. For $N=256$, $\phi_{ac}$ is corrected
for the incompletely relaxed dangling chains to obtain an improved
extrapolation. $m$ is the average RBO of chains between junctions
with 4 active connections to the network. The confinement length $N_{c}$
is is determined in Figure \ref{fig:Average-Segmental-order} from
the average RBOs as function of $k$. $z$ is a single parameter fit
of this data using equation (\ref{eq:mkrev}). }
\end{table}

The poly-disperse samples were obtained from mono-disperse melts of
256 chains of 512 monomers using the same density, lattice size, and
boundary conditions as above. The melt was relaxed for $10^{9}$ simulation
steps, which is about five reptation times \cite{key-23}. Afterward
either 7876 or 1732 monomers were inserted into the melt to model
two-functional cross-links. The mixtures were relaxed for another
reptation time of the chains without crosslinking. Ten conformations
equally separated in simulation time were selected as starting point
for the cross-linking reactions. Here, cross-linking was allowed for
any monomer of the chains under three additional restrictions as in
a previous work \cite{key-16,key-62}: each chain monomer can undergo
only one reaction, the number of reactions is limited to two for the
cross-links, and two consecutive monomers along one chain are not
allowed to react with the same cross-linker. For each of the 20 samples,
a degree of 100\% cross-linking was achieved rapidly. The number average
strand lengths in the sample is approximately $N\approx8$ and 32
monomers. The number average length of active strands is about $8$
and 35 monomers while the weight average active strand length is about
$14$ and 74 monomers respectively. Since the weight average strand
length determines the sample average quantity of active segments,
the samples are characterized below by using the weight average number
of segments per active strand.

For data analysis, network conformations were recorded at a sampling
time interval $\Delta t$ of $10^{5}$ MCS for a period of $10^{8}$
MCS (all networks with average strand length $N\le128$) and $2\times10^{8}$
simulation steps for all other samples. This increased the accuracy
of the entangled data for $N=256$ as compared to our previous study
\cite{key-22}. Diffusion of the sample was corrected by subtracting
the motion of the center of mass for the analysis of monomer fluctuations.
RBOs were computed from time averaged monomer positions.

\section{Ideal networks: affine and phantom network model\label{sec:Theory-and-Simulation}}

\subsection{Theory}

Notation: square brackets $\left[\right]$ are used in the sections
below to denote ensemble averages, angle brackets $\left\langle \right\rangle $
display time averages, bold letters like $\mathbf{R}$ represent vectors
and double arrows $\overleftrightarrow{S}$ indicate tensors. All
distribution functions are normalized such that the integral over
all events is one.

\begin{figure}
\begin{center}\includegraphics[width=0.3\columnwidth]{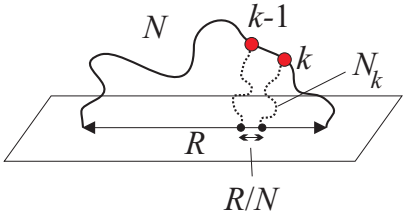}\end{center}

\caption{\label{fig:Affine}Affine network model. A polymer chain of $N$ segments
(black curve) is fixed at both ends to the non-fluctuating elastic
background (plane at bottom) at distance $R$. Fluctuations of the
monomers (red spheres) around the time averaged monomer positions
(black dots) can be modeled by virtual chains $N_{k}$ (dotted lines).}
\end{figure}

Let us assume a Gaussian distribution 
\begin{equation}
P(N,\mathbf{R})=\left(\frac{3}{2\pi Nb^{2}}\right)^{3/2}\exp\left(-\frac{3\mathbf{R}^{2}}{2Nb^{2}}\right)\label{eq:PRN}
\end{equation}
for the probability density of the end-to-end vector $\mathbf{R}$
of a chain made of $N$ segments, each with a root mean square length
$b$ in three dimensional space. We further assume that there are
no correlations among neighboring segments (freely jointed chain).
Finally, we label the segments by $k$ from 1 to $N$ and move the
origin of the coordinate system to the beginning of segment 1.

For the following discussion we use the concept of virtual chains
to describe the time average monomer fluctuations in space. Consider
a monomer that is part of a network of Gaussian chains. Due to network
elasticity, the monomer fluctuates around a time average position
in space. Because of the Gaussian nature of the chains, this fluctuation
can be described by a harmonic potential or, alternatively, by the
distribution of the end-to-end vectors of a ``virtual'' Gaussian
chain of a particular number of segments that is fixed at the average
monomer position in space. Note that below all monomer fluctuations
are expressed by the length of the corresponding virtual chains to
simplify the analysis.

\subsubsection{Affine network model}

In the \emph{affine network model}, ideal chains are assumed to be
attached to the so-called ``non-fluctuating elastic background''
\cite{key-14} that deforms affinely with macroscopic deformations
of the sample, see Figure \ref{fig:Affine}. The long time average
of all $N$ segments is given by the straight line connecting the
fixed ends of the chain, $\mathbf{R}$. In this limit, all time averaged
segments have the same orientation and length due to the constant
force acting along the chain. Therefore, we obtain for the time average
vector of any segment $k$ 
\begin{equation}
\left\langle \mathbf{b}_{k}\right\rangle =\mathbf{R}/N.\label{eq:bk}
\end{equation}

The fluctuations of monomer $k=1,...,N-1$ attached to Segments $k-1$
and $k$ can be computed by solving the Rouse matrix of the network
chain. As a result one finds \cite{key-11} that the time average
monomer fluctuations are equivalent to the fluctuations in space of
the endpoint of a virtual chain of 
\begin{equation}
N_{k}=k(N-k)/N\label{eq:nk}
\end{equation}
monomers that is attached with its other end at position $\left\langle \mathbf{R}_{k}\right\rangle =k\mathbf{R}/N$
to the non-fluctuating elastic background. Therefore, the time average
root mean square fluctuations $\sqrt{\left\langle \Delta R(k)^{2}\right\rangle }$
of monomer $k$ around its average position $\left\langle \mathbf{R}_{k}\right\rangle $
can be described by a virtual chain of $N_{k}$ segments, 
\begin{equation}
\sqrt{\left\langle \Delta R(k)^{2}\right\rangle }=\left\langle \left(\mathbf{R}_{k}(t)-\left\langle \mathbf{R}_{k}\right\rangle \right)^{2}\right\rangle ^{1/2}=bN_{k}^{1/2}.\label{eq:DRaf}
\end{equation}

The residual bond orientation $m_{k}$ of segment $k$ (RBO, also
called the ``vector order parameter'' \cite{key-73,key-60}) can
be defined via an auto-correlation function of the bond orientations,
whereby averages need to be taken over a large number of initial conformations,
see ref. \cite{key-60}. $m_{k}$ can be computed also from considering
the time average position of the monomers at the ends of Segment $k$
\begin{equation}
m_{k}=\left(\left\langle \mathbf{R}_{k}\right\rangle -\left\langle \mathbf{R}_{k-1}\right\rangle \right)^{2}/b^{2}.\label{eq:mk2}
\end{equation}
Note that equation (\ref{eq:mk2}) is the most precise way of measuring
$m_{k}$ from a given set of conformations and it is mathematically
equivalent with considering any available conformation as initial
conformation for the averaging of the auto-correlation functions.

The time average conformation of an ideal chain with fixed ends in
the limit of an infinitely long sampling interval $\Delta t\rightarrow\infty$
is given by the straight line connecting both chain ends, cf. Figure
\ref{fig:Affine}. Because of equation (\ref{eq:bk}) we obtain for
this limit

\begin{equation}
m_{k}=\frac{\left\langle \mathbf{b}_{k}\right\rangle ^{2}}{b^{2}}=\frac{\left\langle \mathbf{R}^{2}\right\rangle }{b^{2}N^{2}}.\label{eq:mk}
\end{equation}

Let us now discuss mono-disperse networks with uniform strand length
$N$ between the junctions. If we insert the ensemble average of the
Gaussian distribution, $[\mathbf{R}^{2}]=b^{2}N$, into equation (\ref{eq:mk}),
we find for the ensemble average of the RBO 
\begin{equation}
[m_{k}]=N^{-1}\label{eq:m}
\end{equation}
independent of $k$. The distribution of RBOs is, therefore, fully
determined by the length distribution of end-to-end vectors of the
$N$-mers 
\begin{equation}
P(N,R)4\pi R^{2}\mbox{d}R=4\pi R^{2}\left(\frac{3}{2\pi Nb^{2}}\right)^{3/2}\exp\left(-\frac{3R^{2}}{2Nb^{2}}\right)\mbox{d}R.\label{eq:PR}
\end{equation}
Substituting $m=R^{2}/(b^{2}N^{2})$ we can transform this probability
density into a probability density of RBOs for Gaussian chains (the
so-called ``Gamma distribution''): 
\begin{equation}
P(N,m)2\pi m^{1/2}b^{3}N^{3}\mbox{d}m=N^{3/2}\sqrt{\frac{27m}{2\pi}}\exp\left(-\frac{3mN}{2}\right)\mbox{d}m.\label{eq:PM}
\end{equation}

This distribution was used in previous works to fit experiment and
simulation data. The agreement between equation (\ref{eq:PM}) and
the data was often poor, in particular for dry networks \cite{key-60}.

\begin{figure}
\begin{center}\includegraphics[width=0.33\columnwidth]{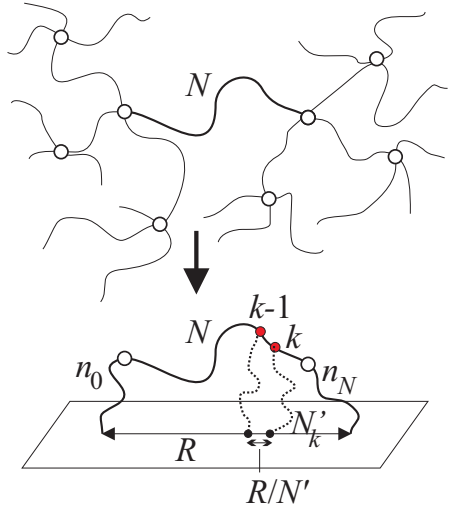}\end{center}

\caption{\label{fig:Phantommodell}Reduction of ideal phantom network (top)
to the corresponding affine chain (bottom, cf. Figure \ref{fig:Affine}).
The virtual chains $n_{0}$ and $n_{N}$ model the fluctuations of
the cross-links (white spheres), if the $N$-mer in between is removed.}
\end{figure}

It has been shown \cite{key-60,key-98,key-19} for Gaussian chains
with fixed ends that the residual coupling constant $s_{k}$ as obtained
in multi-quantum NMR experiments is related to the RBO via 
\begin{equation}
s_{k}=\frac{3}{5}m_{k}.\label{eq:skm-1}
\end{equation}
Moreover, the scaling predictions for $s_{k}$ and $m_{k}$ as function
of time are identical in the entangled regime (cf. section \ref{sec:Time-averageing-and}).
Therefore, $s_{k}\propto m_{k}$ for non-entangled \emph{and} entangled
chains and all scaling laws derived for the RBO apply also for the
experimental data of  dry networks.

\subsubsection{Phantom network model}

In the \emph{phantom network model,} one assumes that restrictions
to junction fluctuations result exclusively from the network connectivity
\cite{key-14,key-13,key-61}. In general, this would require to solve
the Rouse matrix of the entire network, which is impossible for networks
of unknown connectivity (as in experiments) and can be done only numerically
for randomly connected networks (simulations). Therefore, we use the
standard approximation of an ideally branching network structure and
the resulting predictions for the phantom network model to derive
an analytical expression. In general, the phantom model description
for a given network chain can be reduced \cite{key-14} to a combined
chain of 
\begin{equation}
N'=n_{0}+N+n_{N}\label{eq:N'}
\end{equation}
segments, for which the virtual chains $n_{0}$ and $n_{N}$ model
the cross-link fluctuations at both ends of the $N$-mer, after the
$N$-mer has been removed from the network \cite{key-14}. The combined
chain $N'$ is attached with its ends to the non-fluctuating elastic
background. Therefore, we can rewrite all results of the affine model
using $k'=n_{0}+k$ instead of $k$ and $N'$ instead of $N$. To
distinguish between affine and phantom model in the equations we indicate
below all quantities related to the phantom model by an apostrophe.
For instance, the monomer fluctuations in space are described by virtual
chains $N_{k}'=k'(N'-k')/N'$. These modifications lead to 
\begin{equation}
[m]=N'^{-1}=\frac{f-2}{f}N^{-1},\label{eq:m2}
\end{equation}
whereby the right equation only holds for the case of a perfect \cite{key-49}
$f$-functional network of mono-disperse chains $N$ \cite{key-11}.
In the phantom model (as in the affine model), the average residual
RBO is proportional to the inverse of the elastic strand $N'$. Note
that network defects will show a clear effect on the local distribution
of $m$ in the framework of the phantom model, since any defect will
lead to a local variation of the cross-link fluctuation and $N'$.
In order to reduce this additional degree of complexity, we restrict
the data analysis below to strands that are connected at both ends
to cross-links with four active connections to the network. As can
be deduced from the derivation of the above right equation in ref
\cite{key-11}, this procedure removes the largest corrections due
to imperfect network structure for networks with a low average degree
of imperfection.

\subsection{Comparison with simulation data}

\begin{figure}
\begin{center}\includegraphics[angle=270,width=0.7\columnwidth]{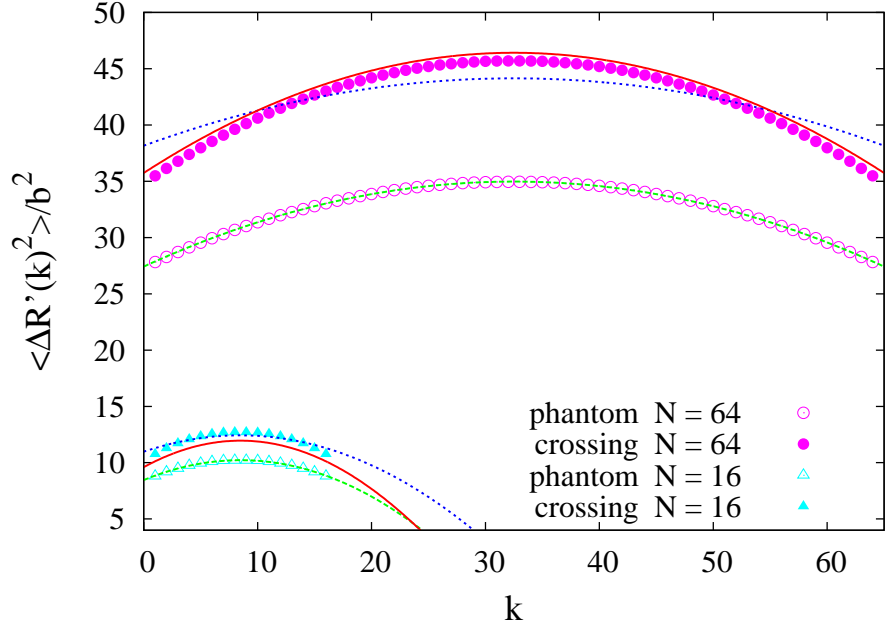}\end{center}

\caption{\label{fig:Monomer-fluctuation}Mean square monomer fluctuation as
function of monomer position $k$ along the chain for phantom and
crossing networks of different $N$. Dashed green line and dotted
blue lines: fit with phantom model. Red continuous line: phantom model
prediction combined with excluded volume correction.}
\end{figure}

\textcolor{black}{For the phantom and crossing networks having }\textcolor{black}{\emph{N}}\textcolor{black}{{}
= 16 and 64, Figure }\textcolor{black}{\emph{\ref{fig:Monomer-fluctuation}}}\textcolor{black}{{}
shows plots of the mean square monomer fluctuations against the position
k along the chain}\textcolor{black}{\emph{.}} The data are normalized
by the root mean square bond length $b^{2}$. Thus, the ordinate in
Figure \ref{fig:Monomer-fluctuation} displays the number of segments
$N_{k}'$ in the corresponding virtual chain that models the fluctuations
of monomer $k$. For the phantom networks, we find excellent agreement
for the $k$-dependence when the length of the virtual chains $n_{0}$
and $n_{N}$ was adjusted to fit the data. The length of the virtual
chains, however, shows some interesting finite size effect, which
is discussed in section \ref{sub:Finite-size-monomer}.

Another interesting observation is that the data of the ``crossing''
networks show a different $k$-dependence as predicted by the phantom
network model in Figure \ref{fig:Monomer-fluctuation}. This is visible
in the different curvature of simulation data and theoretical prediction.
In particular, we find clearly enhanced time average fluctuations
of the monomers in networks with excluded volume, even though the
dynamics is slowed down as compared to the phantom simulations. This
behavior indicates that a lower extensional force is required to stretch
the chains with excluded volume. The different curvature of the simulation
data and the phantom model prediction further indicates that the force
is not lower by a constant factor, instead, the effective force reduction
must be a function of the internal distance along the chains.

Both observations are in line with previously observed corrections
to ideal chain conformations in dense melts \cite{key-42} that arise
from the incompletely screened excluded volume at short distances
along the chains. This is commonly associated with the bare excluded
volume of the monomers \cite{key-24}, but was also identified with
the bulk modulus effectively measured for those systems \cite{key-17}.
In order to test whether this effect accounts for the differences
between the phantom and the crossing data, we perform the following
test: the fit results of the phantom simulations are used as input
for the virtual chain lengths. The phantom prediction for fluctuations
is corrected as described at equation (10) in ref \cite{key-17} by
a factor of 
\begin{equation}
\frac{\left\langle \Delta R(k)^{2}\right\rangle }{\left\langle \Delta R'(k)^{2}\right\rangle }\approx\frac{b_{e}^{2}}{b^{2}}\left(1-\frac{c_{e}}{\sqrt{N_{k}'}}\frac{b_{e}}{b}\right)\label{eq:r/r}
\end{equation}
with parameters $b_{e}\approx1.23b$ and $c_{e}\approx0.41$ as reported
in \cite{key-17}. The result is shown in Figure \ref{fig:Monomer-fluctuation}
as continuous red line. We observe good agreement with the simulation
data, when considering that the phantom and crossing data are collected
from two independent simulation runs of two networks with same connectivity
but with and without excluded volume interactions respectively. In
consequence, Figure \ref{fig:Monomer-fluctuation} leads to the proposal
that the estimate for the phantom contribution to network modulus
should be corrected for real networks with respect to the effect of
excluded volume, since the screening of excluded volume does not depend
on entanglements \cite{key-17}. Furthermore, we use equation (\ref{eq:r/r})
below to correct the effect of incompletely screened excluded volume
for the entangled samples.

\begin{figure}
\begin{center}\includegraphics[angle=270,width=0.7\columnwidth]{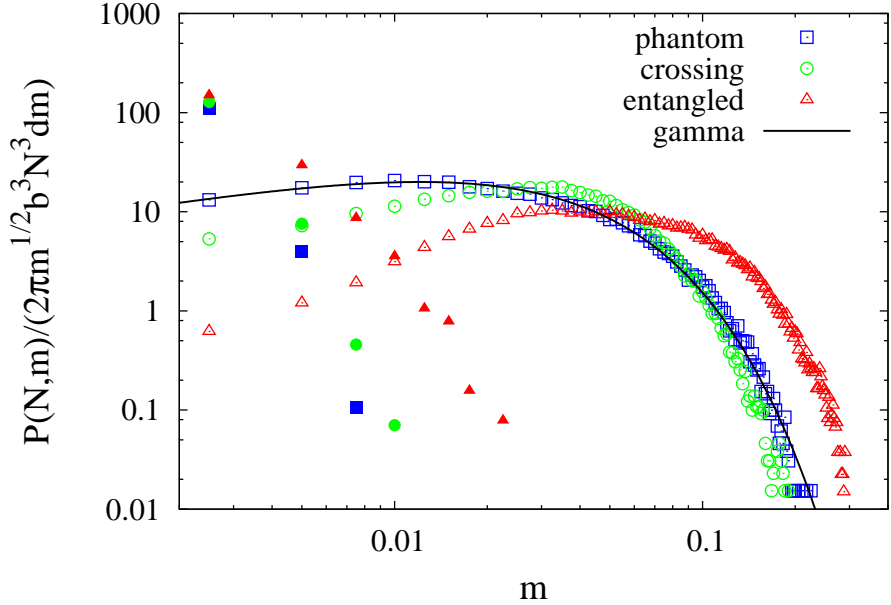}\end{center}

\caption{\label{fig:Distribution-of-order}Distribution $P(m)$ of the RBO
for phantom networks, crossing networks, and regular networks of $N=16$.
The gamma distribution at equation (\ref{eq:PM}) was fitted with
variable $N'$ to the phantom data. Open symbols show results of active
network chains between two junctions that connect 4 active strands.
Full symbols show the results of dangling chains. }
\end{figure}

The distribution of the RBO $P(m)$ of equation (\ref{eq:PM}) is
compared with simulation data of the shortest networks $N=16$ at
Figure \ref{fig:Distribution-of-order}. A fit with a variable chain
length $N'$ as only parameter yields an apparent $N'=27.5\pm0.1$.
The monomer fluctuation data, however, would suggest a clearly larger
$N'\approx17+2\cdot n_{0}\approx41$ (see section \ref{sub:Finite-size-monomer}
for $n_{0}$). This discrepancy can be removed when considering that
the ratio $R^{2}/N^{2}$ enters in computing the RBO, cf. equation
(\ref{eq:mk}). We find that the average square end-to-end vectors
are enlarged by approximately $40\%$ as compared to the ideal prediction
$R^{2}=b^{2}N$, which is roughly in agreement with equation (\ref{eq:r/r}).
Note that the phantom networks were originally entangled networks
for which the excluded volume was switched off. Apparently, the enlarged
state of the chains at crosslinking is conserved by the fixation of
the network via the periodic boundary conditions. Taking into account
this enlarged chain size, we find $N'\approx39$ from the distribution
function of RBOs. This result is in reasonable agreement to the $N'\approx41$
as obtained from the analysis of monomer fluctuations.

Figure \ref{fig:Distribution-of-order} also shows that there are
only small differences concerning the distribution of RBOs for ``phantom''
networks as compared to ``crossing'' networks. The latter show a
slightly enlarged average RBO with a weak depletion at small $m$.
This depletion can be explained by the soft repulsion of connected
active junctions via the excluded volume of all monomers attached
to these junctions. The data of entangled networks in Figure \ref{fig:Distribution-of-order}
is clearly shifted compared to both non-entangled model networks.
Also the dangling material of the ``network'' data shows a distinct
delay concerning relaxation as compared to non-entangled models. Both
observations are remarkable, since strands of 16 monomers are clearly
shorter than previous estimates for the entanglement length, which
amounted in 30 monomers or larger \cite{key-23,key-20}. The clear
shift of the dangling chain data indicates that a similar analysis
of experimental data can separate inactive parts of the network, if
a possible peak at smallest RBOs is excluded.

\begin{figure}
\begin{center}\includegraphics[angle=270,width=0.7\columnwidth]{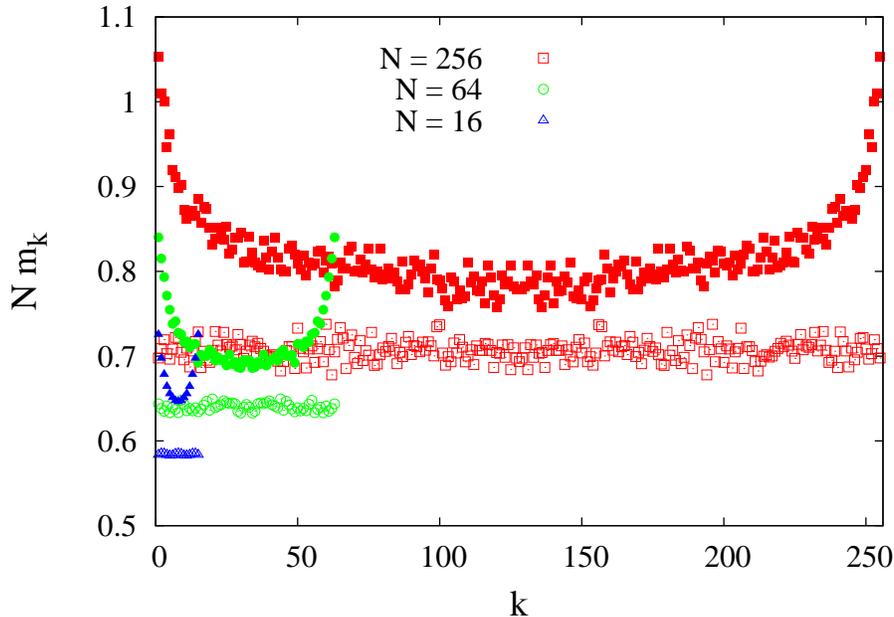}\end{center}

\caption{\label{fig:Average-vector-order}$m_{k}$ as function of the position
$k$ along the chain. Results are multiplied by $N$. Open symbols
show data of phantom networks, filled symbols data of crossing networks.}
\end{figure}

The RBO is averaged for segments of same $k$ and plotted as function
of $k$ in Figure \ref{fig:Average-vector-order} for the ``crossing''
and ``phantom'' networks. The simulation data of the ``phantom''
networks shows that the RBO is independent of $k$ as predicted by
equation (\ref{eq:mk}). The phantom model predicts for a perfectly
branching (no finite loops) network with weight average active functionality
$f_{a}$ of the junctions that 
\begin{equation}
m_{k}N=N'^{-1}N\approx(f_{a}-2)/f_{a}.\label{eq:nk-1}
\end{equation}
As discussed in section \ref{sub:Finite-size-monomer}, $f_{a}$ in
the networks is increasing with $N$ and also, the average chain size
grows as function of $N$ due to preparation conditions (entangled
network with excluded volume switched off after crosslinking) as discussed
above. This explains the increasing $Nm_{k}$ for larger $N$.

For the ``crossing'' simulations, we find an enhancement of RBOs
close to chain ends in the range of $15-30\%$ compared to the middle
of the chains. This enhancement is attributed to the enlarged excluded
volume interactions near the cross-links. Interestingly, the RBO remains
larger as the phantom prediction in the middle of the chains. A close
inspection of the simulation data revealed that the time average monomer
positions of an active chain are not exactly (beyond statistical error)
following a straight line between the two cross-links attached to
both ends. We argue that overlapping chains are competing for the
same lattice positions. This leads to a weak distortion of the average
monomer positions as compared to the phantom networks. Therefore,
an enlarged RBO is detected.

In summary of this section we find excellent agreement between the
``phantom'' networks and the predictions of the phantom model. The
``crossing'' networks show additional effects of excluded volume,
which lead to increased monomer fluctuations and weakly enhanced RBOs,
in particular close to chain ends. Therefore, the phantom model can
be used as a first order approximation to develop a model for the
average segment orientations of entangled networks. Corrections due
to excluded volume for monomer fluctuations will be introduced below
where necessary.

\section{Entangled networks: average segment orientations\label{sub:Entangled-networks:-average}}

\begin{figure}
\begin{center}\includegraphics[width=0.39\columnwidth]{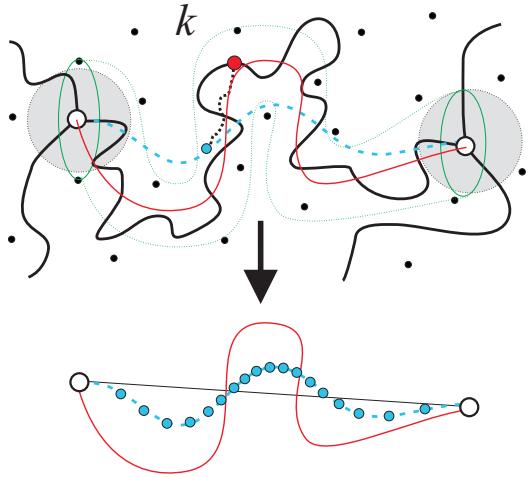}\end{center}

\caption{\label{fig:Average-segment-positions}Upper part: Black dots show
schematically entanglements with active strands that dynamically fluctuate
in space with time. White spheres show cross-links, the gray volume
represents a typical fluctuation volume of a crosslink. Black lines
show polymer strands, the imaginary surrounding time average tube
of the chain in the middle is sketched by the green lines, while the
primitive path of the tube is given by the red line. Monomer $k$
(red sphere) fluctuates mainly along this curved tube leading to an
average monomer position (blue sphere) that is not located on the
primitive path. Lower part: Time average positions of all monomers
(blue spheres) are compared with primitive path (red line) and end-to-end
vector of the chain.}
\end{figure}

The problem that is analyzed in this section is depicted in Figure
\ref{fig:Average-segment-positions}: entanglements confine the motion
of monomer $k$ to a small region in space that might be shaped like
a tube. Since this ``tube'' is not perfectly straight and monomer
$k$ can move a large distance along this curved object, the average
monomer positions neither fall on the primitive path (the center line
of the confining potentials) nor onto the straight line connecting
the cross-link positions as in the affine or phantom model (cf bottom
of Fig. \ref{fig:Average-segment-positions}). The RBOs, however,
are related to the vector connecting the time average monomer positions.
We shall see that the RBO reflects the contour of the primitive path
as averaged by the motion of the monomers along the confining tube.
How this can be analyzed from computer simulations and experimental
data are discussed in the present section.

\subsection{Theory}

For our derivation, we use three assumptions that are consistent with
a slip-tube model for rubber elasticity \cite{key-14}: 
\begin{enumerate}
\item The motion of the monomers is confined to a tube like region in space.
\item The primitive path of the tube performs a random walk in space.
\item The segments of the chains can slide along the contour of the tube,
whereby the motion \emph{along} the chain is limited by network connectivity
(as described by the phantom model), while the \emph{perpendicular}
motion is limited by entanglements.
\end{enumerate}
In\emph{ entangled polymer networks} with a degree of polymerization
$N$ of the network strands larger than the entanglement degree of
polymerization $N>N_{e}$, the monomer and cross-link fluctuations
are additionally constrained as compared to phantom networks by entanglements
with neighboring strands. In the framework of the tube model, this
is typically expressed by a confining potential that restricts perpendicular
fluctuations of the polymers to a tube like region in space \cite{key-40},
see Figure \ref{fig:Average-segment-positions}. In analogy to a stretched
polymer \cite{key-24}, we use $N_{c}$ to describe the diameter of
the tube 
\begin{equation}
a\approx bN_{c}^{1/2}\label{eq:tube diameter}
\end{equation}
(cf. the introduction). Within the \emph{frame of reference of the
tube} (primitive path), the chains have an average end-to-end distance
\cite{key-24} given by 
\begin{equation}
\left[L\right]\approx aN/N_{c}\approx bN/N_{c}^{1/2}.\label{eq:L}
\end{equation}

Let us consider first an entirely straight tube and use the index
\emph{t} to denote the frame of reference of the tube. For simplicity,
we use the phantom model for a chain of $N'=N+n_{0}+n_{N}$ monomers
to describe the motion of the monomers along the tube (which is not
confined by the tube potential) by a virtual chain length $N'_{k}$,
see equation (\ref{eq:nk}). The ends of the combined chain $N'$
are, thus, attached to the elastically active background at $x_{0,t}=0$
and $x_{N',t}=\left[L\right]N'/N$ in order to satisfy equation (\ref{eq:L}).
The probability of finding monomer $k$ at positions $x_{t}$ along
the tube is computed by the one-dimensional contribution \cite{key-11}
of the three dimensional Gaussian distribution 
\begin{equation}
P_{t}(k,x_{t})\approx\sqrt{\frac{3}{2\pi b^{2}N'_{k}}}\exp\left(-\frac{3\left(x_{t}-\left\langle x_{k,t}\right\rangle \right)^{2}}{2b^{2}N'_{k}}\right),\label{eq:11}
\end{equation}
whereby $\left\langle x_{k,t}\right\rangle =\left[L\right](n_{0}+k)/N'$
is the average position of monomer $k$ in the coordinate system of
the tube.

It is typically assumed that the tube performs a random walk in space
\cite{key-24} whereby the step lengths cannot exceed the tube diameter
for maintaining random walk statistics of the full chain. The fluctuations
of inner monomers of entangled chains with $N'_{k}\gg N_{\mbox{\emph{p}}}$
are dominated by fluctuations along the tube \cite{key-11}, while
the motion of monomers with $N_{k}\lesssim N_{p}$ remains essentially
isotropic. The reptation like motion of monomer $k$ along the tube
is limited by the virtual chain length $N'_{k}$. Therefore, for long
chains with $N'_{k}\gg N_{p}$, the time average root mean square
fluctuations of monomer $k$ \emph{in space} are approximately 
\begin{equation}
\sqrt{\left\langle \Delta\mathbf{R}_{en}(k)^{2}\right\rangle }\approx bN_{p}^{1/4}N'{}_{k}^{1/4}.\label{eq:10}
\end{equation}
It is important to point out that in the limit of very long entangled
chains, $N'_{k}\gg N_{p}$, the monomer fluctuations are only related
to the correlation length of the primitive path and not to the tube
diameter, since the contribution of the fluctuations perpendicular
to the tube axis become ignorable. Note that the above relation is
used to estimate $N_{p}$ from monomer fluctuations.

Along a \emph{perfectly straight} \emph{tube}, the RBO of an entangled
chain stretched to $\left[L\right]$ adopts a value of $m=1/N_{c}$
and the RBO is proportional to stress as mentioned in the previous
discussion about the affine model. Bonds that fluctuate along a \emph{curved
tube} of length $\left[L\right]$ sample tube sections of different
orientations. The sampling of different tube orientations reduces
the average RBO of the segments. In order to obtain an analytical
solution, the orientation correlation between the tube sections at
position $x_{t}$ and at the average monomer position $\left\langle x_{k,t}\right\rangle $
along the tube is approximated by an exponential decay for the tube
orientation $O_{t}(x_{t})$ 
\begin{equation}
O_{t}(x_{t})\approx\exp\left(-|x_{t}-\left\langle x_{k,t}\right\rangle |/c\right),\label{eq:12}
\end{equation}
with decay length $c\approx bN_{p}^{1/2}/2$. The resulting reduction
in RBOs from $1/N_{c}$ due to tube curvature is obtained by integrating
$\int_{-\infty}^{\infty}P_{t}(k,x_{t})O_{t}(x_{t})dx_{t}$. This yields
\begin{equation}
m(N,k)\approx\frac{1}{N_{c}}\cdot e^{y^{2}}\cdot\mbox{erfc}\left(y\right)\label{eq:131}
\end{equation}
with the complementary error function \cite{key-99} $\mbox{erfc}(y)$
and 
\begin{equation}
y^{2}=2N'_{k}/3N_{p}.\label{eq:y}
\end{equation}
We expect $N_{c}\propto N_{p}\propto N_{e}$ in the limit of $N_{k}\rightarrow\infty$.
Thus, one obtains as asymptotic result 
\begin{equation}
m\propto N^{-1/2},\label{eq:14}
\end{equation}
for the center monomers in long chains. Since the different RBO near
the chain ends leads to a contribution of order $N^{-1}$, we expect
that equation (\ref{eq:14}) holds for average RBO of active network
strands.

The result of equation (\ref{eq:14}) is clearly different from the
predictions of the affine or phantom network model with $m\propto N^{-1}$
or $m\propto N^{-1}+N_{e}^{-1}$ that both are obtained under the
assumption \cite{key-5,key-97} that the RBO must be proportional
to the modulus of the network, $m\propto G$.

\subsection{Comparison with simulations and experiment\label{sub:Comparison-with-simulations}}

\begin{figure}
\begin{center}\includegraphics[angle=270,width=0.7\columnwidth]{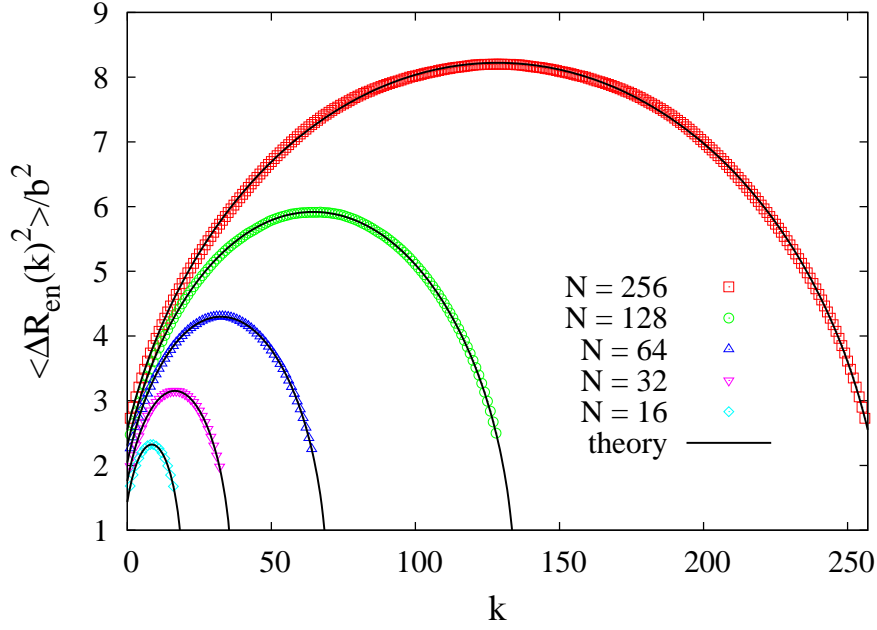}\end{center}

\caption{\label{fig:Monomer-fluctuations-in}Mean square monomer fluctuations
in entangled networks as function of monomer position $k$ along the
chain normalized by the square bond length $b^{2}$. The theoretical
lines are fits of equation (\ref{eq:10}) with $N_{p}$ as single
adjustable parameter, since the crosslink fluctuations as given by
$n_{0}$ can be determined independently. The results for $N_{p}$
are given in Table \ref{tab:mono-1}.}
\end{figure}

Similar to the previous section, only active network strands connecting
pairs of cross-links with 4 active connections are analyzed. This
restriction is necessary, since crosslinks with a smaller number of
active strands show larger fluctuations at the chain ends.

The $N'{}_{k}$ dependence of eq. (\ref{eq:10}) reproduces well the
monomer fluctuations as function of $k$ in all samples. The cross-link
fluctuations as approximated by the virtual chain length $n=n_{0},n_{N}$
(cf. Table \ref{tab:mono-1}) grow clearly weaker than the phantom
prediction $\sim N^{1/2}$, remain smaller than $N_{c}$, but do not
yet fully saturate for large $N$. This is to be expected, since active
cross-links join $\ge3$ active chains. Thus, the fluctuations of
active cross-links in an entangled network must be limited to a volume
$V\lesssim a^{3}.$ The fit parameter $bN_{p}^{1/4}$ seems to converge
towards 4.96 for large $N$. The reduction of this parameter for small
$N$ is possibly the result of the beginning transition to the non-entangled
regime and could be related to a tighter trapping of active chains
for small $N$. Note that the mean square monomer fluctuations of
the crossing and phantom networks are much larger (cf. Figure \ref{fig:Monomer-fluctuation})
and follow a different $N'_{k}$ dependence.

The good agreement between simulation data and theory in Figure \ref{fig:Monomer-fluctuations-in}
is rather surprising. Apparently, there are two opposite corrections
that largely cancel each other. First, as discussed around equation
(\ref{eq:r/r}), the incompletely screened excluded volume interactions
enlarge monomer fluctuations. This can be modeled by an effective
bond length $b_{e}$ with $b_{e}^{2}/b^{2}\approx1.5$ for the bond
fluctuation model at polymer volume fraction $\phi=0.5$ \cite{key-17}.
According to the results of the crossing networks above, we expect
similarly \emph{increased} apparent fluctuations in networks (since
the end monomers are connected to several chains, the apparent $b_{e}^{2}/b^{2}$
is expected to be $\gtrsim1.5$). Second, we know that $n<N_{c}$
and thus, the monomers attached to the cross-links fluctuate rather
isotropically, while fluctuations of inner monomers $N_{k}\gg N_{p}$
are dominated by the fluctuations along the tube. In this limit, only
one directional component contributes to fluctuations. Thus, we expect
that the measured square fluctuations in the coordinate system along
the tube for chains $N\gg N_{p}$ converge to 1/3 of the unconstrained
isotropic fluctuations. This leads to a \emph{reduction} of the measured
square fluctuations in Figure \ref{fig:Monomer-fluctuations-in} of
approximately $3^{1/2}\approx1.7$ for very long $N$. Both corrections
converge as function of $N'{}_{k}^{1/2}$ towards the limiting behavior
for inner monomers for large $N$. According to the above numerical
estimate, most of these corrections cancels each other. Since, we
also do not observe significant differences between theory and simulation
data, we conclude that $N_{p}$ can be calculated via equation (\ref{eq:10})
without the need of additional corrections or a numerical constant.
However, for a different simulation method with a clearly different
$b_{e}/b$, a correction could become necessary.

\begin{figure}
\begin{center}\includegraphics[angle=270,width=0.7\columnwidth]{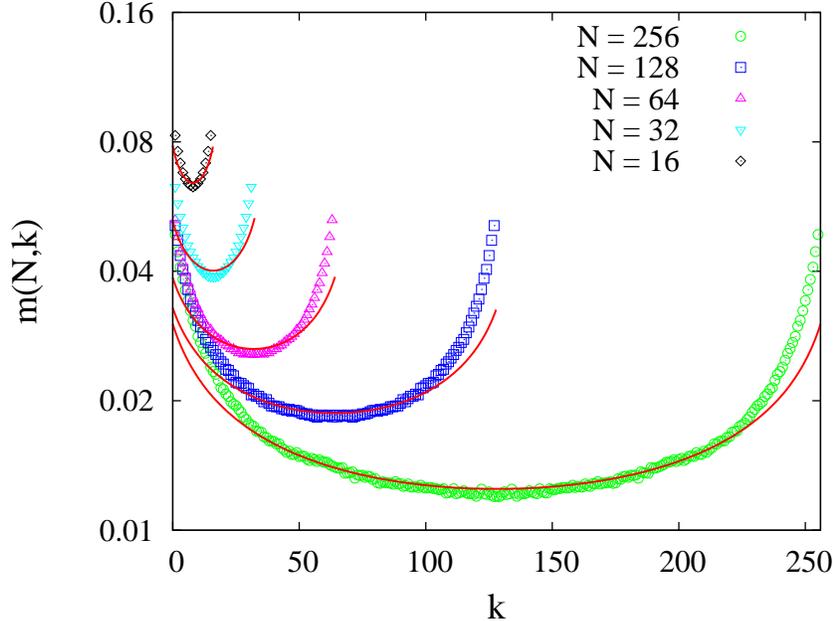}\end{center}

\caption{\label{fig:Average-Segmental-order}Average RBO as function of monomer
position $k$ along the chains. Theoretical lines are fits with equation
(\ref{eq:131}) with $N_{c}$ as parameter and including the corrections
for enlarged chain size and fluctuations as described in the text.
The results for the fit of $N_{c}$ are given in table \ref{tab:mono-1}.}
\end{figure}

\begin{figure}
\begin{center}\includegraphics[angle=270,width=0.7\columnwidth]{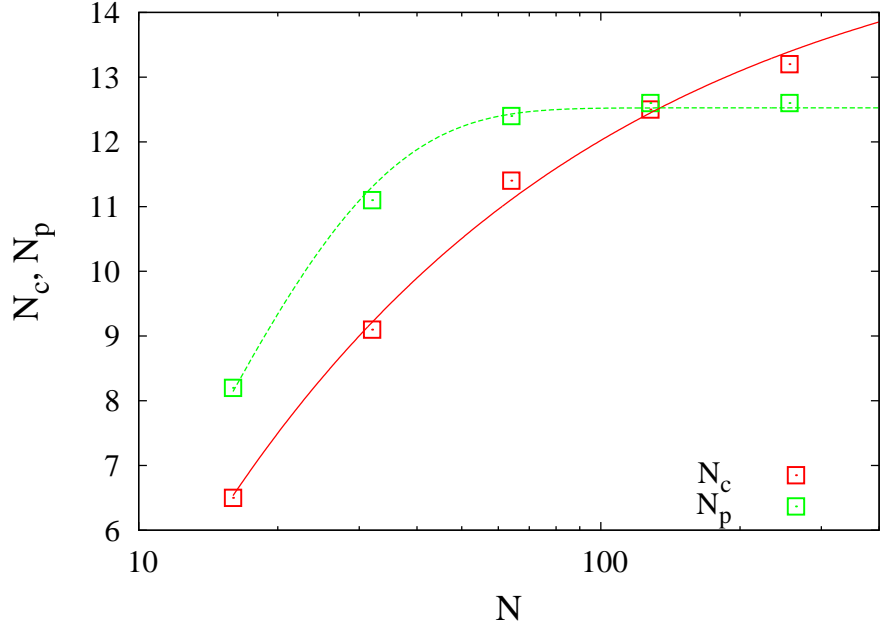}\end{center}

\caption{\label{fig:-and-}$N_{c}$ and $N_{p}$ as function of $N$. The lines
are the approximations for the convergence towards the asymptotic
limit, equation (\ref{eq:Np}) and (\ref{Nc}), as discussed qualitatively
in the introduction.}
\end{figure}

The RBO $m(N,k)$ is analyzed as function of bond index $k$, see
Fig. \ref{fig:Average-Segmental-order}. We use the same subset of
active chains and parameters $n$ and $N_{p}$ as in Figure \ref{fig:Monomer-fluctuations-in}.
As discussed for the crossing networks in the previous section, the
partially expanded chain conformations (due to the incompletely screened
excluded volume) need to be corrected. This leads to a) an enlargement
of the average square chain extension, and thus, we obtain an additional
factor of approximately $b_{e}^{2}/b^{2}\left(1-c_{e}b_{e}/(b\sqrt{N}\right)$
for predicting the RBO in equation (\ref{eq:131}). b) As shown in
the previous chapter, monomer fluctuations along the tube, as given
by $N'_{k}$, are enlarged by $b_{e}^{2}/b^{2}\left(1-c_{e}b_{e}/(b\sqrt{N'_{k}}\right)$.
This can be corrected for in equation (\ref{eq:y}). c) $N_{p}$ requires
no modification according to the discussion of the preceding section.
Equation (\ref{eq:131}) including these corrections is fitted to
the innermost 80\% of the data of each network in Figure \ref{fig:Average-Segmental-order}
using $N_{c}$ as the \emph{only} adjustable parameter. The data near
the ends is discarded due to the extra contribution from excluded
volume as shown in Figure \ref{fig:Average-vector-order}. The result
for $N_{c}$ is summarized in Table \ref{tab:mono-1} and both $N_{c}$
and $N_{p}$ are plotted in Figure \ref{fig:-and-}. We find that
both $N_{p}$ and $N_{c}$ are increasing and converge for $N\rightarrow\infty$
to the asymptotic limit as proposed in the introduction. Figure \ref{fig:-and-}
indicates that $N_{c,\infty}/N_{p,\infty}\approx5/4$, which agrees
with the result $N_{p}\le N_{c}$ of Helfand et al. \cite{key-27,key-74}.

In our previous work \cite{key-22} we did not distinguish between
$N_{p}$ and $N_{c}$ and omitted the above corrections for fluctuations
and expansion in chain size. As a result, a fit of the data suggested
a decreasing de-correlation length for larger $N$. The present analysis
shows, that this observation was caused by the neglect of the above
corrections for chain size and monomer fluctuations.

\begin{figure}
\begin{center}\includegraphics[angle=270,width=0.7\columnwidth]{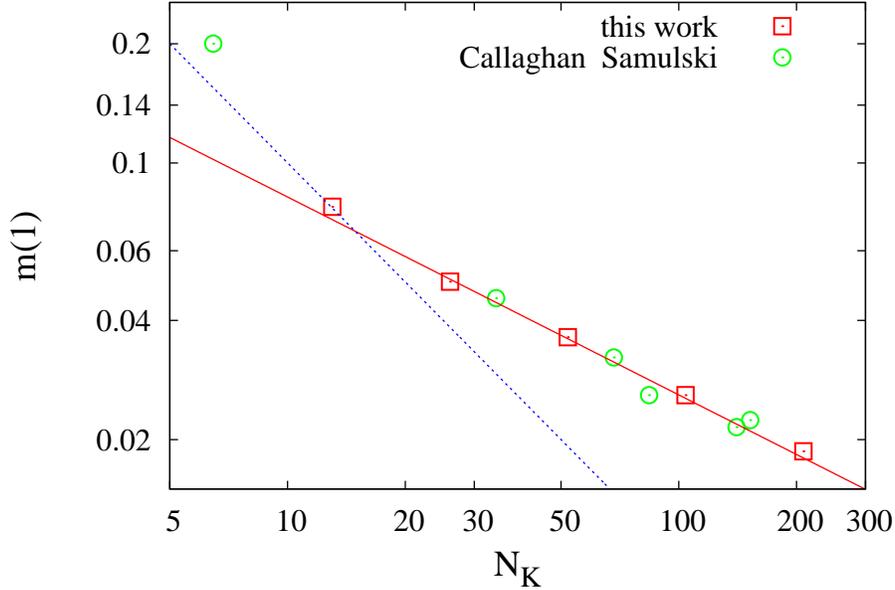}\end{center}

\caption{\label{fig:order of N}Average RBO of active chains as function of
the number of Kuhn segments $N_{K}$ extrapolated to networks without
defects. The continuous red lines correspond to a fit with power law
$m\propto N^{-1/2}$ while the blue dotted line shows $m=N^{-1}$.
The experimental data are taken from Ref \cite{key-6} and normalized
as described in the text.}
\end{figure}

The average RBO $m$ is analyzed as function of the number of Kuhn
monomers $N_{K}$ in Figure \ref{fig:order of N} and compared with
the experimental data of table 1 of Ref \cite{key-6}. In order to
superimpose both data sets, we compute the number of Kuhn segments
$N_{K}=Nb/b_{e}=N/1.23$ with $b_{e}$ of Ref \cite{key-17} for the
simulation data and use the molar mass of a Kuhn Monomer for PDMS
$M_{0}=381g/mol$ from Ref \cite{key-11} for the experimental data.
Next, we extrapolate the average RBO to networks without defects by
computing $m(1)=m/\phi_{ac}^{2/3}$ for both simulations and experiment.
Finally, we shift the experimental data for the residual moment along
the ordinate (here: by a factor of $1/30000000$) to optimize the
overlap of the data in the large $N_{K}$ region. This is to test,
whether there is a \emph{qualitative} agreement between simulations
and experiment for entangled networks. This comparison is possible,
since in bulk materials, the residual moment is $\propto s\propto m$
as discussed in Ref \cite{key-60}. Note that the perdeuterated sample
of Ref \cite{key-6} is left out from the plot (similar to the analysis
of Ref \cite{key-6}) due to its different relaxation behavior. If
we ignore the two data points for the smallest chain lengths in Figure
\ref{fig:order of N}, we find that both simulation and experimental
data of Ref \cite{key-6} follow the predicted scaling of $N^{-1/2}$
over about one decade in $N_{K}$. This observation is in contrast
to the proposal of Refs \cite{key-5,key-97}, which related the RBO
to the equivalent length $N_{el}$ of an elastic chain describing
the network elasticity in the framework of the affine model: 
\begin{equation}
m\propto\frac{1}{N_{el}}=c\left(\frac{1}{N}+\frac{1}{N_{e}}\right)\label{eq:m_vergleich}
\end{equation}
using a normalization constant $c$. Note that in Ref \cite{key-6}
it was attempted to use equation (\ref{eq:m_vergleich}) to describe
the RBO as function of $N$. But this lead to a different power for
$m$ as function of $N_{el}$ as proposed in equation (\ref{eq:m_vergleich}).

Let us now discuss the \emph{experimental} data in more detail. The
location of the cross-over between the entangled and the non-entangled
regimes may be estimated by assuming that the data point at smallest
$N_{K}$ is clearly in the $N^{-1}$ regime of non-entangled networks.
If we use this data point to fix the $N^{-1}$ regime, the crossover
to the $N^{-1/2}$ is around $22$ Kuhn monomers. Since $N_{e}\approx32$
for PDMS \cite{key-11}, we estimate $N_{p}\approx2N_{e}/3$. If we
assume that $N_{p}$ is of the order of the average weight between
two slip links, this result is close to the prediction $N_{p}/N_{e}=4/7$
of Ref \cite{key-14} or near the estimates $1/2$ of Ref \cite{key-55}
and $4/5$ of Ref \cite{key-56}. This agreement supports that our
model may be used for a refined quantitative analysis of entangled
polymers.

\begin{figure}
\begin{center}\includegraphics[angle=270,width=0.7\columnwidth]{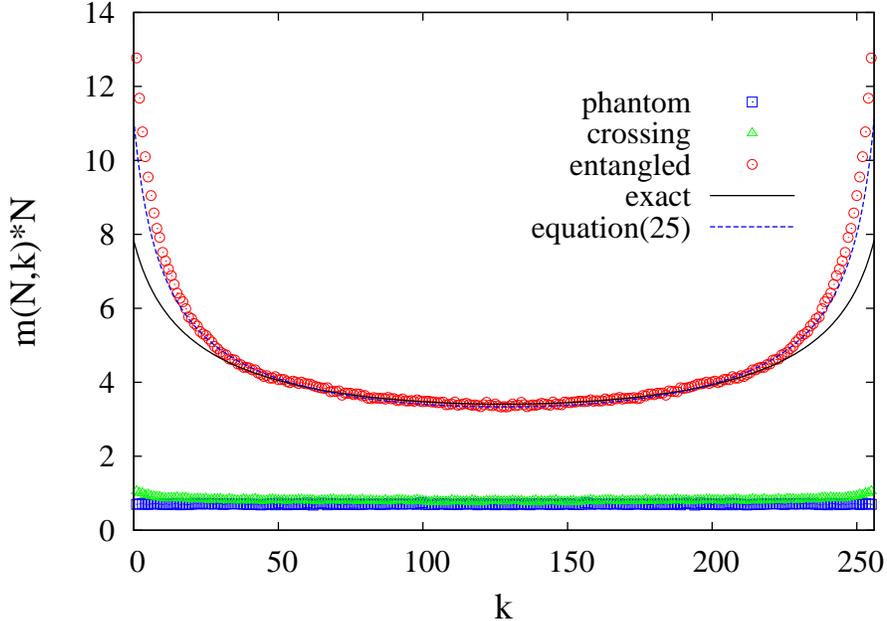}\end{center}

\caption{\label{fig:Comparison-of-scaling}Average RBOs for all mono-disperse
networks with $N=256$ multiplied by $N$. Comparison of equation
(\ref{eq:mkrev}) and exact solution (equation (\ref{eq:131})) with
the data of entangled networks (red circles). For comparison, the
data of the non-entangled two model samples is also included.}
\end{figure}

In Fig \ref{fig:Comparison-of-scaling} we compare data of the phantom,
crossing, and entangled networks with $N=256$ in order to show the
clear qualitative and quantitative differences for the entangled networks.
In particular, the strong increase of the RBO towards the chain ends
is characteristic for entangled networks. Note that this behavior
is supported by experimental data of Gronski \emph{et al}. \cite{key-46,key-64},
who determine a clearly enhanced RBO close to cross-links as compared
to inner parts of the chains.

Let us now try to find a simplified approximation for the RBO as function
of $k$. We expect $m(k)$ to be reduced by a factor proportional
to the return probability to the most probable tube section $\sim\left(N_{p}/N'_{k}\right)^{1/2}$
as compared to the the straight tube result $1/N_{c}$, since all
neighboring uncorrelated tube sections are averaged to zero. Furthermore,
we ignore the corrections for increased chain size and enlarged fluctuations.
For the remaining approximation, we introduce a numerical constant
$z$ to fit the data in Fig. \ref{fig:Average-Segmental-order} by
\begin{equation}
m_{k}\approx\frac{z}{\sqrt{N_{p}N'_{k}}}.\label{eq:mkrev}
\end{equation}
This approximation works surprisingly well as shown at Fig. \ref{fig:Comparison-of-scaling}
and even compensates some deviations near the chain ends. The resulting
$z$ vary less than $15\%$ over more than a decade of $N$ (see table
\ref{tab:mono-1}). Note that we still require $N_{p}$ and the virtual
chain length $n$ of table \ref{tab:mono-1} as parameters that are
independently determined from monomer fluctuations. We have to point
out that any difference between $N_{c}$ and $N_{p}$, all corrections
due to the extended chain conformations, enlarged monomer fluctuations,
the numerical coefficients of equation (\ref{eq:131}) and (\ref{eq:y})
and all effects of the crosslinking process are summed up in $z$.
A quantitative analysis of the RBO, therefore, requires the computation
of the more detailed approach above. Nevertheless, we think that equation
(\ref{eq:mkrev}) is quite useful in practice, since it allows for
a) a quick prediction of the average RBO as function of $k$ once
$z$ has been determined for a particular $N$ at a particular polymer
volume fraction (for a given polymer or simulation model). Furthermore,
we use equation (\ref{eq:mkrev}) as a simplification to derive the
distribution of RBOs in the following sections.

\section{The distribution of residual bond orientations in entangled networks\label{sub:Distribution-of-order}}

\subsection{Theory}

In the model developed above, we computed the average RBO as function
of $k$ by assuming an average tube length $\left[L\right]$ and an
average tube curvature as expressed by an exponential de-correlation
function. This is a reasonable approximation for deriving the average
RBO, but it requires further generalization for computing the distribution
of the RBOs.

First, we have to clarify, that the approximation below focuses on
intermediate chain lengths $N\gtrsim N_{p}$ as it is typical for
polymer networks. For very long chains, any variation of the local
tube curvature is averaged out (except of near the ends), since the
monomers will visit a large number of different tube sections $\propto\left(N_{k}/N_{p}\right)^{1/2}\gg1$
while moving back and forth along the tube. Also, the fluctuations
in the tube length grow only $\propto\left(N/N_{p}\right)^{1/2}$
and thus, become unimportant for $N\rightarrow\infty$. Apparently,
both corrections lead to a qualitatively similar broadening of the
distribution of RBOs as function of $N$. Therefore, we discuss below
only the effect of tube length fluctuations that is better documented
in literature. We expect that the effect of fluctuations in the local
tube curvature leads to an additional broadening of the RBO distribution.

Let us assume that cross-linking occurs rather instantaneously and
that the relaxation of the network chains during cross-linking does
not lead to a an equilibration of two connected tubes until the full
network structure is fixed. Under these conditions, the frozen length
distribution of the network tubes is roughly equal to the instantaneous
length distribution of the ensemble of tubes inside the melt prior
to cross-linking.

In the Doi-Edwards model \cite{key-24}, polymers in melts are considered
as Rouse chains inside a tube, whereby the chain is stretched by a
potential of entropic origin for keeping its extension at time average
length $\left\langle L\right\rangle =[L]=bN/N_{c}^{1/2}$. We assume
that the tube length distribution of the melt is frozen in the network.
Assuming a quadratic approximation \cite{key-26} for the free energy
\begin{equation}
F(L)=\frac{3\gamma kT}{2Nb^{2}}\left(L-\left[L\right]\right)^{2}\label{eq:fl}
\end{equation}
with an effective spring constant $\gamma$ of order unity we obtain
normal distributed tube lengths \cite{key-11} around $\left\langle L\right\rangle $
as described by 
\begin{equation}
P(N,L)\mbox{d}L=\sqrt{\frac{3\gamma}{2\pi Nb^{2}}}\exp\left(-\frac{3\gamma(L-\left[L\right])^{2}}{2Nb^{2}}\right)\mbox{d}L.\label{eq:pnle}
\end{equation}
The magnitude of the apparent spring constant could be modified in
a network due to the impact of the local tube curvature, or due to
stress equilibration and redistribution of monomers when connecting
two tubes at their ends.

Comparing equation (\ref{eq:mkrev}) with the result of a straight
tube $m_{k}=1/N_{c}$ we can interpret the fluctuations along the
curved tube as a length contraction of a straight tube by a factor
$\beta=z^{1/2}\left(N_{p}/N'_{k}\right)^{1/4}$, since $m\sim L^{2}$
in this case at constant $N$ and $b$ analogous to the affine model
result $m\sim R^{2}$ at constant $N$ and $b$ (see previous chapter
for parameter $z$). This leads to a contracted apparent tube length
$\beta\left[L\right]$ and a modified spring constant of this ``contracted''
ensemble of chains $\gamma_{c}=\gamma\beta^{2}/z^{2}$. Thus, the
monomer fluctuations along the tube let the distribution of the primitive
path lengths appear contracted \cite{key-67} for the analysis of
the RBO: 
\begin{equation}
P(L,k)\mbox{d}L=\sqrt{\frac{3\gamma_{c}}{2\pi N'b^{2}}}\exp\left(-\frac{3\gamma_{c}(L-\beta\left[L\right])^{2}}{2N'b^{2}}\right)\mbox{d}L.\label{eq:PL}
\end{equation}
Equation (\ref{eq:PL}) can be transformed using the substitution
$m=L^{2}/(b^{2}N'^{2})$ to obtain the distribution of RBOs in entangled
networks 
\begin{equation}
P(m,k)\frac{bN'}{2m^{1/2}}\mbox{d}m=\sqrt{\frac{3\gamma_{c}N'}{8\pi m}}\exp\left(-\frac{3\gamma_{c}N'}{2}\left(m^{1/2}-\beta/N_{p}^{1/2}\right)^{2}\right)\mbox{d}m.\label{eq:pmke}
\end{equation}
This equation allows us to compute the distribution of RBOs for any
labeled monomer $k$ and thus, for the full chain via the summation
\begin{equation}
P(m,N)\mbox{d}m=\sum_{k=0}^{N}P(m,k)\frac{bN'}{2m^{1/2}}\mbox{d}m/(N+1).\label{eq:pme}
\end{equation}
Note that this summation runs over all $N+1$ segments between the
cross-links, while the length distribution was determined by the combined
chain $N'$.

\subsection{Comparison with simulation data}

\begin{figure}
\begin{center}\includegraphics[angle=270,width=0.7\columnwidth]{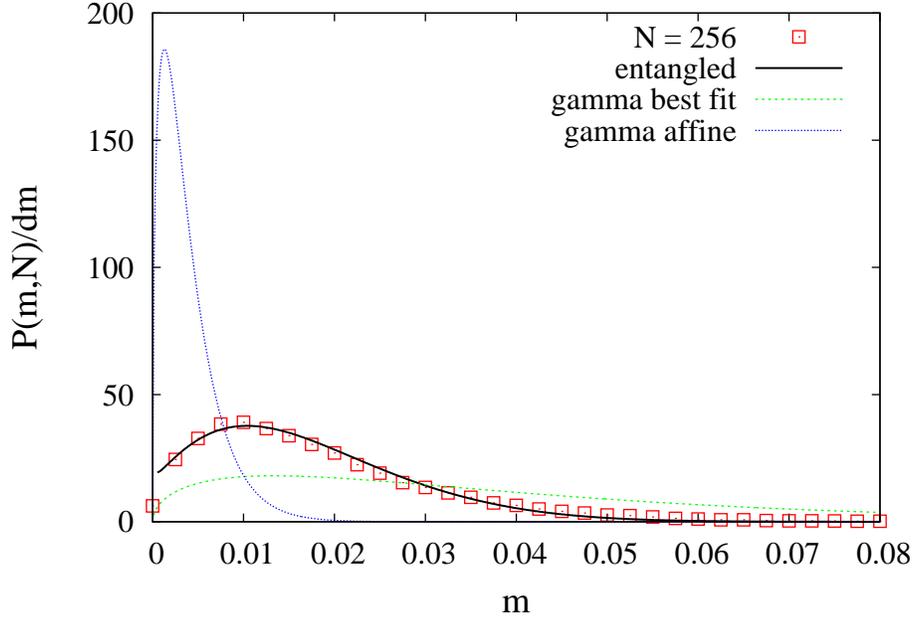}\end{center}

\caption{\label{fig:Order-parameter-distribution}Distribution of RBOs in the
entangled network $N=256$ (data points). The gamma distribution of
equation (\ref{eq:PM}) is plotted using $N=256$ (affine model) or
fitted to the data yielding $N=25\pm5$. The ``entangled'' line
is computed using equation (\ref{eq:pme}) with $\gamma=0.45\pm0.05$
and the previously determined data of table \ref{tab:mono-1} for
$N_{p}$, $z$, and $n$.}
\end{figure}

The distribution of RBOs in the entangled network $N=256$ is plotted
in Fig. \ref{fig:Order-parameter-distribution}. Neither the affine
model (or the phantom model, not shown) nor a best fit using the gamma
distribution with $N$ as adjustable parameter can even \emph{qualitatively}
describe the data. A best fit of the gamma distribution of equation
(\ref{eq:PM}) yields $N\approx25$, which is clearly below $N=256$
and not in agreement with the $N_{p}$ as determined from monomer
fluctuations. A fit of equation (\ref{eq:pme}) using a variable $\gamma$
leads to a clearly better qualitative description of the simulation
data as compared to the gamma distribution.

\begin{figure}
\begin{center}\includegraphics[angle=270,width=0.7\columnwidth]{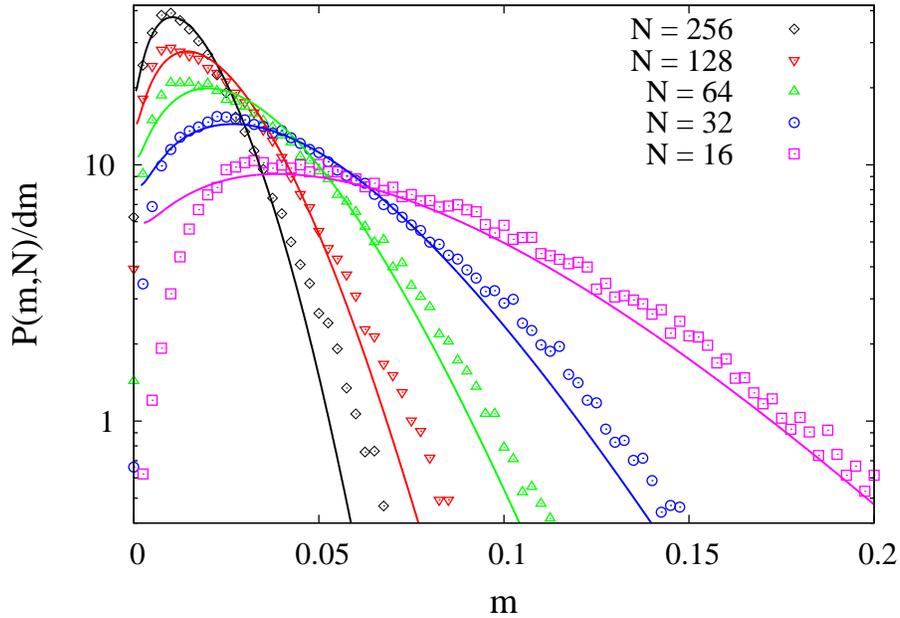}\end{center}

\caption{\label{fig:Distribution-of-residual}Distribution of RBOs in all entangled
networks (data points). Full lines are computed using equation (\ref{eq:pme})
with $\gamma=0.45$ and the data of table \ref{tab:mono-1} for $N_{p}$,
$z$, and $n$.}
\end{figure}

The data of all entangled networks of the present study is fitted
with equation (\ref{eq:pme}) at Figure \ref{fig:Distribution-of-residual}
using a single spring constant $\gamma$ for all networks. In particular
for large $N$ we find good agreement. For small $N$ we observe an
additional depletion at small $m$. We attribute this difference to
the effect of excluded volume: Since the cross-links are attached
to four chains, the cross-links repel each other similarly to the
centers of star polymers in a melt of stars. This removes extremely
short tube lengths from the time averaged conformations and leads
to the depletion at $m\approx0$. Since the average tube length grows
$\propto N$, the effect of this short distance correction vanishes
for $N\rightarrow\infty$.

In the derivation above, we adopted a harmonic potential to derive
the tube length distribution as introduced in the Doi-Edwards model
for entangled polymers. The parameter $\gamma$ as front factor of
this potential is considered to be close to 0.8 \cite{key-11} as
indicated by experimental data. Here, a good agreement is found at
a rather constant $\gamma\approx0.45$, which was used to compute
all solid lines in Figure \ref{fig:Distribution-of-residual}. If
we accept the experimental $\gamma$ as reference, the reduced $\gamma$
of the present study indicates that the RBO distribution is additionally
broadened, possibly by local fluctuations in tube curvature: in contrast
to melts where diffusing chains continuously create and sample new
tube sections, the fixed network chains sample only a particular set
of tube sections and curvatures. This may explain our observations,
but requires more detailed investigations to proof this claim.

\section{Poly-disperse Networks\label{sec:Polydisperse-Networks}}

\begin{figure}
\begin{center}\includegraphics[angle=270,width=0.7\columnwidth]{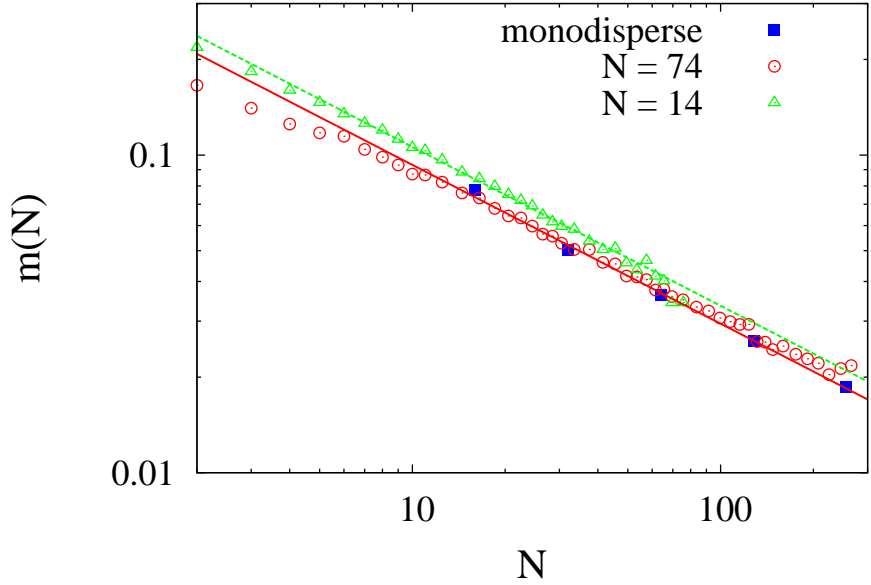}\end{center}

\caption{\label{fig:Estimation-of-the}Average RBO as function of strand length
$N$ in cross-linked networks. The lines are fits with $N^{-1/2}$.}
\end{figure}

Following the discussion in section \ref{sub:Distribution-of-order},
we expect that the RBO parameter $m(N)$ of a strand of $N$ monomers
in poly-disperse samples follow $m(N)\approx z/(NN_{p})^{1/2}$ for
sufficiently large $N>N_{p}$. The data in Figure \ref{fig:Estimation-of-the}
clearly supports the predicted scaling. Note that all data points
have been extrapolated to a density of 100\% active material, since
the weight fraction of dangling chains leads to a significant shift
of the data. The data of the networks $N=14$ shows slightly enhanced
order, probably due to a tighter trapping of the strands at the crossover
to the $N^{-1}$ dependence of non-entangled networks, since $N_{w,ac}\approx N_{p}$.
For comparison, Figure \ref{fig:Estimation-of-the} additionally includes
a data set of the mono-disperse series of entangled networks. The
agreement with the data of the mono-disperse networks implies that
poly-disperse networks can be modeled by a superposition of mono-disperse
networks.

This observation is used to compute the distribution of the RBOs in
randomly crosslinked networks. The contribution of a particular active
strand length $N$ is computed according to equation (\ref{eq:pme})
and weighted by the $w_{N,ac}$ of the $N$-mer in the sample. The
number fraction $n_{N}$ of strand lengths can be approximated by
$n_{N}\approx\exp\left(-N/N_{n}\right)$ \cite{key-16,key-43}, whereby
the number average strand length $N_{n}$ is computed from the number
density of crosslinks. Accordingly, the weight fraction of active
strands of $N$ monomers is approximately (assuming no significant
difference in the distributions of active and inactive strands) $w_{N,ac}\approx Nn_{N}/\sum_{N}Nn_{N}$.
The distribution function of RBOs is then approximated by 
\begin{equation}
P_{c}(m)\mbox{d}m\approx\sum_{N=1}^{\infty}w_{N,ac}P(m,N)\mbox{d}m.\label{eq:PCM}
\end{equation}

The numerical solution to this equation is computed using the average
virtual chain lengths $n=3.6$ and $5.3$, which were determined from
the cross-link fluctuations in the networks (after normalizing by
$f/(f-1)$ to account for the definition of the virtual chains). Furthermore,
the same $z$ as for the corresponding end-linked samples with $N=64$
was used for comparison.

\begin{figure}
\begin{center}\includegraphics[angle=270,width=0.7\columnwidth]{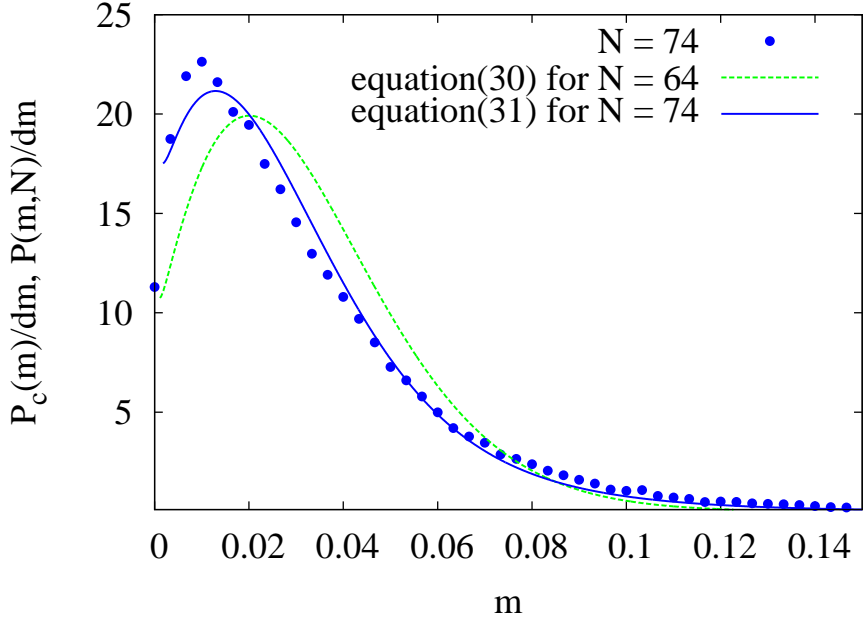}\end{center}

\caption{\label{fig:Comparison-of-the}The distribution of RBOs of cross-linked
networks compared with the predictions for end-linked (dashed line)
and crosslinked (continuous line) networks.}
\end{figure}

The data of the cross-linked sample $N=74$ is shown in Figure \ref{fig:Comparison-of-the}
and compared with a fit of equation (\ref{eq:PCM}) using a variable
$\gamma$ to compute of $P(m)$. The agreement between simulation
data and equation (\ref{eq:PCM}) is good and points towards a slightly
reduced $\gamma\approx0.33$ as compared to the end-linked networks.
Note that the randomly crosslinked networks have been linked rather
instantaneously, while the end-linking reaction was rather slow in
comparison. The different $\gamma$ for these two types of networks,
can be explained, therefore, by a possible stress equilibration and
redistribution of monomers when connecting two tubes at their ends
by end-linking. Compared to the prediction of the end-linked networks,
we further observe in Figure \ref{fig:Comparison-of-the} a shift
of the peak position towards smaller RBO and a larger high molecular
weight tail of the distribution. This is qualitatively as expected
from equation (\ref{eq:PCM}).

\section{Discussion\label{sec:Discussion}}

To relate NMR data to the structure of entangled polymer networks,
the central assumption in previous literature is to treat entanglements
as additional cross-links and thus, the contributions of entanglements
and cross-links to the residual dipolar coupling are additive \cite{key-5,key-97}.
Under these assumptions, the effective elastic chain length $N_{el}$
is determined by NMR and the average network strand length $N$ is
calculated from the average residual dipolar coupling using 
\begin{equation}
N_{el}^{-1}=N^{-1}+N_{e}^{-1}\simeq N_{e}^{-1}\,\,\,\,\,\,{\mbox{for}}\,\, N\gg N_{e},\label{eq:1}
\end{equation}
whereby $N_{e}$ denotes the entanglement length.

In the present work, it was shown that this approach is not consonant
with slip-link or slip tube models of entangled networks, see Ref
\cite{key-14}. In conclusion, we cannot consider entangled chains
being glued to each other like cross-linked chains. Instead, the strands
fluctuate along a confining tube-like region in space leading to an
RBO $m\propto N^{-1/2}$. This has important consequences for the
interpretation of experimental data and for the theory of rubber elasticity.

First, the prediction of decreasing RBO $m\sim N^{-1/2}$ is \emph{not}
in contrast with the stress optical rule \cite{key-24}: NMR measures
the \emph{time average} of \emph{individual} \emph{bond orientations},
which then becomes averaged over the entire sample. But stress is
defined by the \emph{ensemble average} of \emph{all instantaneous
forces }at a given time \cite{key-24}. The former is measured for
the same bonds at \emph{different} positions in space, while the latter
is the average over different monomers or bonds that are at the \emph{same}
location in space. Both quantities match coincidentally in the framework
of the affine and phantom network model, since there, time and ensemble
averages are fully exchangeable and stress is solely transmitted along
the network bonds. Entanglements, by definition, confine the fluctuations
of a chain to a given region in space depending on the particular
entanglements of a given chain. Therefore, the time averages of individual
chains are no longer exchangeable with the ensemble averages over
all chains, which must lead to different results for the RBO (or the
tensor order parameter) as compared to stress.

Second, the pronounced dependence of the RBO as function of $k$ (see
Figures \ref{fig:Average-Segmental-order} and \ref{fig:Comparison-of-scaling})
can only be explained by introducing a mechanism that reduces the
RBO towards the inner monomers of the chains. It was shown that a)
monomer fluctuations increase towards the inner monomers of the chains
and b) that these enhanced fluctuations can lead to a reduction of
the RBO, since a larger number of uncorrelated tube sections can be
visited. Such a behavior is only consonant with slip-link or slip
tube models \cite{key-14,key-68,key-69,key-70}, which allow the segments
to slide through slip links or along the tube.

Third, with the present model it is possible to separately analyze
the correlation length of the primitive path $N_{p}$ and the strength
of the confining potential as expressed in the confinement length
$N_{c}$. As discussed in section \ref{sub:Comparison-with-simulations},
a quantitative analysis of NMR data \cite{key-6} suggests that the
de-correlation length of the tube $N_{p}$ is about $N_{p}/N_{e}\approx2/3$,
which is in the range of theoretical estimates \cite{key-14,key-56,key-55}
that predict $N_{p}=N_{e}/2$, $4N_{e}/7$, or $4N_{e}/5$. Best agreement
is found with the model of \cite{key-14} that was used as a basis
for the computations of the current paper.

In literature, several simulation works exist, which are closely related
to the present study. Sotta et al. \cite{key-29,key-47,key-65} use
a ``network'' model of individual chains fixed at their ends similar
to the work of Yong and Higgs \cite{key-36} and analyze the effect
of deformation on the RBO. The missing network structure allows dis-entanglements
near the chains ends, therefore, we do not expect significant impact
of the remaining entanglements for the studies using short chain lengths
\cite{key-47,key-65,key-36}. A later work \cite{key-29} covers larger
$N$, which are comparable to the present study. Interestingly, the
author mentions that the RBO close to chain ends is generally different
from that in the middle of the chains, but a detailed analysis of
this point is missing.

Recently, the group of Prof. Cohen analyzed RBOs in experiments on
end-linked PDMS networks and using computer simulations \cite{key-32,key-34,key-35}.
The experimental results motivated a DE-compositon of the NMR spectra
into a wide and narrow one, whereby it was shown that inelastic chains
contributed only partially to the narrow spectrum. This decomposition
is therefore problematic since the authors cannot provide a striking
physical explanation to justify the source of a second population
of chains. In the present work, the simulation data are fit without
introducing a second distribution of chain lengths or RBOs. In contrast,
to the assumption in \cite{key-32,key-34,key-35} it is shown, that
a different RBO appears along the same chains, making the separation
into two major populations obsolete. However, the results of \cite{key-32,key-34,key-35}
point out that it is essential to distinguish between active and inactive
material. For instance, the simulation data of Fig 6 or Ref. \cite{key-35}
shows nearly identical average RBO for short and long chains in samples
of 60\% mol fraction of short chains. For those samples it is known
that a high fraction of short chains form inactive dangling rings
\cite{key-43,key-44} and thus, we have to expect a large fraction
of chains with RBO close to zero. In consequence, in reference \cite{key-35}
the average segmental orientation becomes a function of composition
and the average RBO of long chains eventually becomes larger as for
short chains.

The determination of the entanglement length is one of the key questions
of recent computational studies on entangled polymers, see Ref. \cite{key-39}
for a review. The methods discussed in Ref \cite{key-39} either require
modifying the samples for analysis (``primitive path analysis'')
or to use a static analysis of conformations. For a most direct comparison
of our work with previous data it is preferable to compare with dynamic
data of unmodified samples. For this comparison, it is important to
recognize that tube length fluctuations (TLF) are largely suppressed
in for entangled active network stands, since the chain ends are tied
to other active chains. But inactive network chains release here a
significant amount of constraints influencing time averaged data.
In the long time limit, this inactive material effectively serves
as a $\theta-$solvent for the active chains dilating the tube to
an apparent $N_{p}=N_{p}(1)\phi_{ac}^{2/3}$ with respect to an $N_{p}(1)$
as determined in a perfect network. Note that for this estimate we
assume that the line density of entangled strands is homogeneously
reduced by the weight fraction of active polymer $\phi_{\mbox{ac}}$.
We further have to point out that all inactive material can be considered
to be relaxed during the time period of the analysis for $N\le128$,
see also section \ref{sec:Time-averageing-and}. For network $N=256$,
however, a separate analysis of the RBOs of dangling chains revealed
that about 25\% of the dangling material is still not fully relaxed
by the end of the simulation run. Therefore, the $\phi_{\mbox{ac}}$
in table \ref{tab:mono-1} was estimated with respect to the fully
relaxed inactive material.

In contrast to networks, constraint release is typically considered
as a less important correction for mono-disperse melts while a dominating
correction is attributed to TLF \cite{key-11,key-24}. For instance,
Kreer \emph{et al.} \cite{key-23} or Paul \emph{et al. }\cite{key-20,key-63}
investigate entangled melts using the simulation method of the present
work at the same lattice occupation of $\phi=0.5$. In ref \cite{key-23},
the relaxation times of melts with $N\le512$ are reported. It is
estimated that $N_{e}\approx38$ based upon the ratio of reptation
times without considering TLF. Recall the discussion at the introduction
and the fact that the reptation times were determined by an analysis
based upon diffusion instead of stress relaxation. Therefore, we expect
that the results of ref \cite{key-23} are more related to the entanglement
spacing and thus $N_{p}$ instead of $N_{e}$. The samples $N_{1}=512$
and $N_{2}=128$ of ref \cite{key-23} are clearly entangled. The
effect of TLF can be estimated by equating the reported ratio for
the relaxation times using the Doi-Edwards correction for TLF for
which we use\cite{key-102} $N_{p}$ instead of $N_{e}$ as variable:
\begin{equation}
\frac{\tau_{rep}(N_{1})}{\tau_{rep}(N_{2})}\approx\frac{N_{1}^{3}[1-\mu\sqrt{N_{p}/N_{1}}]^{2}}{N_{2}^{3}[1-\mu\sqrt{N_{p}/N_{2}}]^{2}}.\label{eq:tn1}
\end{equation}
Solving for $N_{p}$ using the data of ref \cite{key-23} leads to
$N_{p}\approx10\pm1$, whereby the error indicates different results
obtained by considering either end-to-end vector relaxation time $\tau_{ee}$
or the relaxation for the longest mode $\tau_{1}$ while using either
$\mu\approx1$ or a more accurate estimate for $\mu$ from Figure
9.21 of \cite{key-11}. This result is in excellent agreement with
the persistence length of the primitive path $N_{p}$ as determined
in the present work, since the extrapolated data for networks with
out defects yields $N_{p}(\phi_{ac}=1)\approx12$. In fact, following
the discussion in section \ref{sub:Comparison-with-simulations} we
have to conclude that the diffusion data of long entangled chains
must be related to $N_{p}$ instead of $N_{e}$. Note that a similar
shift to $N_{p}=16\pm5$ for the original estimate of $30\pm15$ is
obtained for Ref.\emph{ }\cite{key-20,key-63} when reanalyzing the
delay of 
\begin{equation}
\tau_{rep}/\tau_{Rouse}\approx N[1-\mu\sqrt{N_{p}/N}]^{2}/N_{p}\approx6.67.\label{eq:reprouse}
\end{equation}
Therefore, the data in refs \cite{key-23,key-20,key-63} fully supports
our analysis about monomer fluctuations. But the results for polymer
dynamics may be reinterpreted based upon $N_{p}$ instead of $N_{e}$,
if the model and the analysis of the present work is confirmed by
other studies.

In previous studies on melts of cyclic polymers \cite{key-71,key-72}
using the BFM at $\phi=0.5$, an anti-correlation peak was found in
the correlation function of bonds with a separation distance 11 bonds
along the ring. This observation could not be understood since previous
studies estimate $N_{e}\approx30$ and assume that a de-correlation
of the tube is only possible at $N_{e}$. The data of the present
work, however, finds $N_{p}\approx12$ for ideal networks. The same
result may apply to melts of long rings without further corrections
due to the absence of chain ends. If this agreement is confirmed by
the data of different simulation models or for simulations of chains
with a different stiffness, the analysis of the bond correlations
in melts of long rings could be used for quick determination of the
de-correlation length of the confining tube.

One of the reviewers of this work wondered why a ``strangulation''
of the network chains for $N\rightarrow1$ is not observed. The strangulation
regime was originally discussed \cite{key-100} for unattached chains
that diffuse through a network, whereby the mesh size of the network
becomes smaller than the entanglement length. For the four-functional
neworks of the present study, there is an average of $n\approx8$
strands needed to form a cyclic structure inside the network \cite{key-101}.
The cyclic structures lead to a confinement that can be described
by a virtual chain of length $nN/2$. The non-entangled elasticity
of the network, on the other hand, is described by a chain of $2N$
segments (for ideal $4$-functional network) that is grafted at both
ends \cite{key-14}. Thus, the maximum fluctuations of inner chain
monomers are described by virtual chains of roughly $N<nN/2$ segments.
Since network elasticity leads to a stronger confinement, the network
strands of the present study do not strangulate themselfes when crossing
over to the non-entangled regime. However, the effect of strangulation
can contribute to a measurable enlargement of the residual bond orientation
for $N<N_{e}$, since both strangulation and elasticity have the same
dependence on $N$.

In conclusion, we find supporting simulations and experimental data
for the presented model for the RBO in networks. The data can only
be explained by assuming a chain slippage along a confining tube or
through slip links. Therefore, the results are in agreement with slip
link or slip tube models \cite{key-14} for networks and disagree
with network models that do not explicitly allow a fluctuation of
the monomers along a confining tube.

\section{Summary}

In this work we presented theoretical models and data for average
RBO and the distribution of RBOs for entangled model networks of a)
mono-disperse or b) poly-disperse weight distribution between the
crosslinks, c) non-entangled ``phantom'' model networks and d) non-entangled
``crossing'' model networks with excluded volume interactions.

It is found that the phantom model correctly describes monomer fluctuations
in networks without excluded volume and entanglements. Monomer fluctuations
in networks with excluded volume but no entanglements can be described,
if the phantom model is corrected by the effect of the incompletely
screened excluded volume. This leads to larger monomer fluctuations
and conversely, a weaker force that is necessary to stretch a network
of excluded volume chains as compared to the phantom prediction. This
result is used as a correction for the analysis of the RBOs of entangled
networks.

For the analysis of entangled samples, it is distinguished between
tube confinement, expressed by a confinement length $N_{c}$, and
reorientation of the tube, expressed by a tube de-correlation length
$N_{p}$. $N_{p}$ can be determined from monomer fluctuations of
active chains attached to network junctions that are attached to a
defined number of active strands. The confinement length $N_{c}$
enters in the model by the analogy between tension blobs of a stretched
chain and the stretching of the chains along the primitive path. Both
$N_{p}$ and $N_{c}$ contribute differently to the RBO and thus,
allow a separate analysis using the model of the present work. For
the BFM we find $N_{p}\approx12$ at $\phi=0.5$ and large $N$ when
extrapolating to ideal networks without defects. Previous simulation
data on the dynamics of linear chains \cite{key-20,key-63,key-23}
and cyclic polymers in melts \cite{key-71,key-72} lead to a similar
conclusion after the data on dynamics is corrected for tube length
fluctuations.

The RBO in phantom networks is well predicted by the phantom model
after considering the boundary conditions and the fixation of the
networks in a weakly stretched state. The RBO in crossing networks
with excluded volume but without entanglements shows some weak enhancement
near the chain ends due to increased excluded volume interactions
with attached other chains at the cross-links. The RBOs in entangled
networks are quantitatively and qualitatively different from the non-entangled
networks. First, the average RBO is larger, roughly by a factor $(N/N_{p})^{1/2}$.
Second, there is a pronounced enhancement of RBOs near the chain ends
that is caused by the reduced fluctuations of the monomers near the
crosslinks. The same scaling of the RBO $\sim(NN_{p})^{-1/2}$ is
observed in all entangled networks of the present study and it is
confirmed by experimental data \cite{key-6}. A quantitative analysis
of the RBO of this experimental study suggests that the tube de-correlates
at a degree of polymerization $N_{p}$ of about $2/3$ of $N_{e}$.
This is in good quantitative agreement with the slip tube model \cite{key-14}
and not too far from other estimates \cite{key-55,key-56}. In conclusion,
the $N_{p}$ as determined from the average RBO or the distribution
of the RBOs is shorter than the entanglement length $N_{e}$.

The distribution function of the RBO agrees well with the gamma distribution
in the case of phantom networks. For mono-disperse entangled networks
we obtain a clearly different distribution function that is in good
agreement with the simulation data. The same approach can be used
for randomly cross-linked networks and it also fits to the data. The
key result of our model, that the RBO of entangled networks is $\propto N^{-1/2}$,
does not disagree with the stress optical rule, since NMR does not
measure stress. Only in some particular cases, e.g. of network models
without entanglements, there is an coincidental agreement between
RBO and stress.

\section{Acknowledgment}

The authors thank the computing facilities at the ZIH and IPF Dresden
for a generous grant of computing time. ML thanks J-U. Sommer, S.
Nedelcu, A. Galuschko, K. Saalwächter, E.T. Samulski, M. Rubinstein
and S. Panyukov for helpful discussions. Financial support by the
DFG grant LA 2735/2-1 is acknowledged.

\section{\label{sec:Time-averageing-and}Appendix}

\subsection{Time averaging, relaxation, and comparison with experimental data}

It was shown in a previous paper \cite{key-60} that computing the
residual bond and tensor orientations by using running averages of
the bond and tensor orientations leads to an apparent decay $m\propto t^{-1}$
and $s\propto t^{-1/2}$ respectively. In particular, for completely
uncorrelated bond orientations we have $m=\left(t/\Delta t\right)^{-1}$.
This dependence limits the applicablilty of both averages to regimes
at which the actual residual bond orienation decays not faster than
the runnig averages. Auto-correlation functions as used previously
\cite{key-66} do not suffer from this limitation but require a larger
number of conformations to analyze, since the results are numerically
less stable.

The residual tensor orientation decays $\propto t^{-1/4}$ between
the entanglement relaxation time $\tau_{e}$ and the Rouse relaxation
time $\tau_{R}$ of the strand, $\tau_{e}<t<\tau_{R}$ \cite{key-97}.
The scaling of the RBO is derived in similar manner. In section \ref{sub:Entangled-networks:-average},
we assumed that only the initial tube section is correlated, while
all other tube sections are not correlated. Thus, the RBO is reduced
as function of time proportional to the return probability to the
original tube section $\propto t^{-1/4}$. Since this is the same
scaling as for the residual tensor orientation, we expect that both
residual bond and tensor orientation scale in the same manner as function
of time. Equilibrium is reached at the Rouse time of the virtual chain
describing the fluctuations of the monomers next to the segment. Therefore,
the proportionality $s_{k}\propto m_{k}$ extends from the non-entangled
regime to the entangled regime. In consequence, we can apply all scaling
results for the RBO to the experimental data. A quantiative analysis,
however, requires to show that also the constant of proportionality
is not affected by the crossover from the non-entanged to the entangled
regime.

\subsection{\label{sub:Finite-size-monomer}Finite size monomer fluctuations
in small non-entangled networks}

In this section it is discussed how the finite simulation size affects
the crosslink fluctuations of the non-entangled ``phantom'' and
``crossing'' samples. The entangled samples are not touched by this
effect, since the crosslink fluctuations are strongly suppressed by
entanglements.

For the three phantom networks $N=16,$ 64, and 256 we obtain $n_{0}=11.94\pm0.03$,
$n_{0}=37.44\pm0.05$, and $n_{0}=130\pm1$ when using equation (\ref{eq:DRaf})
and the corresponding average root mean square bond lengths of the
active chains $b=2.68$, $b=2.72$, and $b=2.72$ respectively. The
phantom model \cite{key-11} predicts for ideal $f$-functional networks
without defects 
\begin{equation}
n_{0}=n_{N}=\frac{N}{f-2}.\label{eq:no}
\end{equation}
The networks of the present study have a conversion of approximately
$0.95\%$ and contain small dangling loops and short cyclic structures
of a small number of chains that are less active \cite{key-1,key-9}.
Therefore, we expect that an effective functionality $f_{e}\le3.8=0.95\cdot f$
determines the cross-link fluctuations, whereby $f_{e}$ is expected
to converge towards $3.8$ for increasing $N$, since cyclic defects
vanish for $N\rightarrow\infty$. $f_{e}$ can be estimated in an
approximate manner by solving the above equation for $f$ and using
the fits for $n_{0}=n_{N}$. The above data yields $f_{e}\approx3.34$,
$f_{e}\approx3.74$ and $f_{e}\approx3.97$ respectively when considering
the $N+1$ segments between the junctions.

A detailed analysis of the network structure on the other hand reveals
that the weight average active functionality of the junctions is $f_{a}\approx3.38$,
$f_{a}\approx3.47$ and $f_{a}\approx3.54$ respectively. While the
result for $N=16$ is in good agreement with the fluctuation analysis,
there is a strikingly increasing discrepancy for growing $N$ that
is clearly beyond the error of the data as it is observable in Figure
\ref{fig:Monomer-fluctuation}.

We argue that the reason for the discrepancy between both data sets
is due to two finite size effects. Obviously, subtracting the center
of mass motion leads in small samples to a weak correlation of cross-link
fluctuations (and the material attached to the cross-link) and sample
diffusion and therefore, to the measurement of increasingly smaller
monomer fluctuations with decreasing number of chains. But, the more
pronounced finite size effect originates from having a small network
spanned by periodic boundary conditions: For the networks $N=256$,
the size distribution of the cyclic structures is already dominated
by cycles that are closed through the periodic boundary condition,
see \cite{key-41} for a more detailed discussion. This leads to an
increased coupling of the fluctuations of a cross-link with its own
periodic image. The magnitude of this effect can be estimated by truncating
the recursive computation of monomer fluctuations at a finite number
of generations of chains roughly equal to half the length of an average
``periodic'' network cycle recalling the derivation of the phantom
model (cf. section 7.2.2 of \cite{key-11}). The generation $g$ at
which the total number of cross-links $M/2$ in the structure becomes
comparable to the number of junctions in a tree of average functionality
$f_{a}$ becomes the key quantity for this estimate. It is computed
numerically by solving 
\begin{equation}
1+\sum_{g=1}f_{a}(f_{a}-1)^{g-1}\approx M/2\label{eq:sum}
\end{equation}
for $g$. The solution of this equation leads to a rather small $g\approx5$
for $M=512$ and $f_{a}\approx3.54$. Therefore, this effect leads
to an overestimation of $f_{e}$ of roughly 10\% for $N=256$, while
this overestimation reduces to about 1\% for $N=16$. This qualitative
change in the discrepancy between the numerical data for $f_{e}$
and $f_{a}$ is well reflected by the above data and contributes significantly
to the shift of the RBO in Figure \ref{fig:Average-vector-order}.

\newpage

\section*{Table of Contents Graphic}

\textbf{Monomer Fluctuations and the Distribution of Residual Bond
Orientations in Polymer Networks}

\bigskip

\noindent \emph{Michael Lang}

\bigskip

\bigskip

\bigskip

\begin{center}\includegraphics[width=0.8\columnwidth]{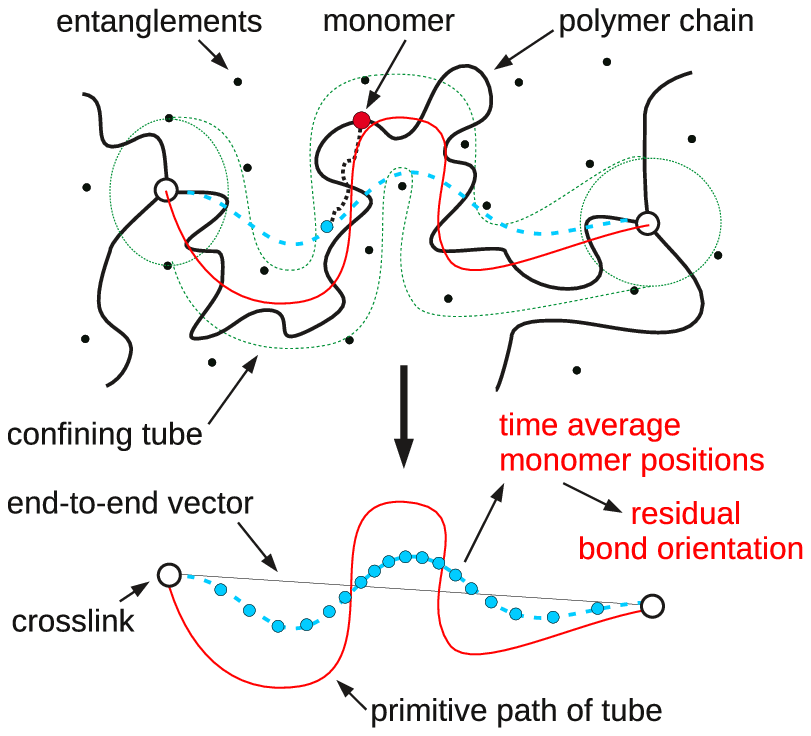}\end{center}

\bigskip


\begin{thebibliography}{References}
\bibitem{key-22}M. Lang, J.-U. Sommer, \emph{PRL} \textbf{104}, 177801
(2010).

\bibitem{key-14}M. Rubinstein, S. Panyukov, \emph{Macromolecules}
\textbf{35}, 6670 (2002).

\bibitem{key-37}M. Rubinstein, \emph{J.Pol.Sci.B} \textbf{48}, 2548
(2010).

\bibitem{key-11}M. Rubinstein, R. Colby, \emph{Polymer Physics,}
Oxford University Press, New York (2003).

\bibitem{key-40}S. F. Edwards, \emph{Proc.Phys.Soc. (London)}\textbf{
92}, 9 (1967).

\bibitem{key-24}M. Doi, S.F. Edwards, \emph{The Theory of Polymer
Dynamics}, Oxford University Press, New York (1986).

\bibitem{key-38}H. Watanabe, \emph{J.Polym.Sci.} \textbf{24}, 1253
(1999).

\bibitem{key-39}S. Shanbhag, M. Kröger, \emph{Macromolecules} \textbf{40},
2897 (2007).

\bibitem{key-52}W. Bisbee, J. Qin, S. T. Milner, \emph{Macromolecules}
\textbf{44}, 8972 (2011).

\bibitem{key-53}R. N. Khaliullin, J. D. Schieber, \emph{Macromolecules}
\textbf{42}, 7504 (2009).

\bibitem{key-54}A. E. Likhtman, Viscoelasticity and Molecular Rheology.
In \emph{Comprehensive Polymer Science}, 2nd ed.; Elsevier, Amsterdam
(2011).

\bibitem{key-73}One of the reviewers pointed out that the ``vector''
or ``tensor order parameters'' are indeed no true order parameters
as used in statistical physics. Therefore, ``residual bond orientation''
or ``residual tensor orientation'' is used in the present work as
replacement.

\bibitem{key-8}K. Saalwächter, \emph{Prog. NMR Spectr. }\textbf{51},
1 (2007).

\bibitem{key-3}K. Saalwächter, P. Ziegler, O. Spyckerelle, B. Haidar,
A. Vidal, J.-U. Sommer, \emph{J. Chem. Phys.} \textbf{199}, 3468 (2003).

\bibitem{key-4}K. Saalwächter, B. Herrero, M.A. Lopez-Manchado, \emph{Macromolecules}
\textbf{38}, 9650 (2005).

\bibitem{key-5}P. Sotta, C. Fülber, D.E. Demco, B. Blümich, H. W.
Spiess, \emph{Macromolecules} \textbf{29}, 6222 (1996).

\bibitem{key-6}P.T. Callaghan, E.T. Samulski, \emph{Macromolecules}
\textbf{33}, 3795 (2000). 

\bibitem{key-59}M. Knörgen, H. Menge, G. Hempel, H. Schneider, M.
E. Ries, \emph{Polymer} \textbf{43}, 4091 (2002).

\bibitem{key-7}P. G. De Gennes, \emph{J. Chem. Phys. }\textbf{55},
572 (1971).

\bibitem{key-56}R. G. Larson, T. Sridhar, L. G. Leal, G. H. McKinley,
A. E. Likhtman, T. C. B. McLeish,\emph{ J. Rheol}, \textbf{47}, 809
(2003).

\bibitem{key-55}R. Everaers, \emph{Phys. Rev. E} \textbf{86}, 022801
(2012).

\bibitem{key-27}E. Helfand, D.S. Pearson, \emph{J. Chem. Phys.} \textbf{79}
2054 (1983).

\bibitem{key-74}M. Rubinstein, E. Helfand, \emph{J.Chem.Phys.} \textbf{79},
2054 (1985).

\bibitem{key-75}Consider an ideal chain that is fixed in space at
both ends and connect one end of a second ideal chain to an arbitrary
monomer of the first chain. The other end of this second chain is
fixed at an arbitrary position in space that allows the second chain
to maintain Gaussian statistics. If the fixed end of the second chain
is not located on the straight line connecting both ends of the first
chain, then, the time averaged conformation of the original chain
is no longer a straight line connecting both ends. Since the time
average conformation is no longer a straight line, the tension along
the chain (as measured by the contour length of the time average conformation)
wass increased by the linking both chains together.

\bibitem{key104}M. Lang, M. Hoffmann, R. Dockhorn, M. Werner, J.-U.
Sommer, JChemPhys 139, 164903 (2013).

\bibitem{key-15}I. Carmesin, K. Kremer, \emph{Macromolecules} \textbf{21},
2819 (1988).

\bibitem{key-16}M. Lang, D. Göritz, S. Kreitmeier, \emph{Macromolecules
}\textbf{38}, 2515 (2005). 

\bibitem{key-62}M. Lang D. Göritz, S. Kreitmeier, \emph{Macromolecules
}\textbf{36}, 4646 (2003).

\bibitem{key-23}T. Kreer, J. Baschnagel, M. Müller, K. Binder, \emph{Macromolecules}
\textbf{34}, 1105 (2001).

\bibitem{key-48}W. Michalke, M. Lang, S. Kreitmeier, D. Göritz, \emph{JChemPhys}.
\textbf{117}, 6300 (2002).

\bibitem{key-42}A. N. Semenov, A. Johner, \emph{Eur.Phys.J}.\emph{E}\textbf{
12}, 469 (2003).

\bibitem{key-60}J.-U. Sommer, K. Saalwächter,\emph{ Eur.Phys.J. E
}\textbf{18}, 167 (2005).



\bibitem{key-98}J.-P. Cohen Addad, \emph{Prog. NMR Spectrosc.} \textbf{20},
1 (1993).

\bibitem{key-19}J.-U. Sommer, W. Chasse, J. L. Valentin, K. Saalwächter,
\emph{Phys.Rev.E} \textbf{78}, 051803 (2008). 

\bibitem{key-13}H. M. James, \emph{J. Chem. Phys.} \textbf{15}, 651
(1947). 

\bibitem{key-61}H. M. James, E. Guth, \emph{J. Chem. Phys.} \textbf{11},
455 (1948).

\bibitem{key-49}i.e. an ideal perfectly branching network without
finite cyclic structures at full conversion.

\bibitem{key-17}J. P. Wittmer, P. Beckrich, H. Meyer, A. Cavallo,
A. Johner, J. Baschnagel, \emph{Phys. Rev. E }\textbf{76}, 011803
(2007).

\bibitem{key-20}W. Paul, K. Binder, D. W. Heermann, K. Kremer, \emph{J.
Phys. II} \textbf{1}, 37 (1991). 

\bibitem{key-99}I. S. Gradshteyn, I.M. Ryzhik, \emph{Table of Integrals,
Series, and Products}, 5th Ed., Academic Press Boston (1994).

\bibitem{key-97}R. C. Ball, P.T. Callaghan, E.T. Samulski, \emph{J.
Chem. Phys.} \textbf{106}, 7352-7361 (1997).

\bibitem{key-46}W. Gronski, D. Emeis, A. Brüderlin, M. M. Jacobi,
R. Stadler, C.D. Eisenbach, Brit.Polym.J 17, 103 (1985). 

\bibitem{key-64}W. Gronski, R. Stadler. M.M. Jacobi, \emph{Macromolecules}
\textbf{17}, 741 (1984).

\bibitem{key-26}M. Doi, N.Y. Kuzuu,\emph{ J.Pol.Sci Pol.Lett }\textbf{18},
775 (1980).

\bibitem{key-67}Note that the contraction is assymmetric and thus,
does not preserve the expectation value of the average RBO. This leads
to a shift of the average RBO of about 5\% for the RBOs of the present
study which is corrected numerically for comparison with the simulation
data.

\bibitem{key-43}J.-U. Sommer, S. Lay, \emph{Macromolecules} \textbf{35},
9832 (2002).

\bibitem{key-68}S. F. Edwards, T. A. Vilgis, \emph{Rep. Prog. Phys.}
\textbf{51}, 243 (1988).

\bibitem{key-69}R. C. Ball, M. Doi, S. F. Edwards, M. Warner, \emph{Polymer}
\textbf{22}, 1010, (1981).

\bibitem{key-70}S. F. Edwards, T. A. Vilgis, \emph{Polymer} \textbf{27},
483 (1986).

\bibitem{key-29}P. Sotta, \emph{Macromolecules} \textbf{31}, 8417
(1998).

\bibitem{key-47}P. Sotta, P.G. Higgs, M. Depner, B. Deloche, \emph{Macromolecules}
\textbf{28}, 7208 (1995). 

\bibitem{key-65}M. Depner, B. Deloche, P. Sotta, \emph{Macromolecules}
\textbf{27}, 5192 (1994).

\bibitem{key-36}C. W. Yong, P. G. Higgs, \emph{Macromolecules} \textbf{32},
5062-5071 (1999).

\bibitem{key-32}G. D. Genesky, T. M. Duncan, C. Cohen, \emph{Macromolecules}
\textbf{42}, 8882 (2009).

\bibitem{key-34}B. M. Aguilera-Mercado, C. Cohen, F. A. Escobedo,
\emph{Macromolecules} \textbf{42}, 8889 (2009).

\bibitem{key-35}B. M. Aguilera-Mercado, G. D. Genesky, T. M. Duncan,
C. Cohen, F. A. Escobedo, \emph{Macromolecules} \textbf{43}, 7173
(2010).

\bibitem{key-44}W. Michalke, M. Lang, A. Buchner, S. Kreitmeier,
D. Göritz, \emph{CompTheoPolSci} \textbf{11} 459 (2001).

\bibitem{key-63}W. Paul, K. Binder, D. W. Heermann, K. Kremer, \emph{J.
Chem. Phys.} \textbf{95}, 7726 (1991).

\bibitem{key-102}The theory for equation (\ref{eq:tn1}) was developed
for $N_{e}$ by assuming that any correlated section of the primitive
path contributes $kT$ to the modulus. It was further estimated that
$\mu\approx1.47$ for a continuous Rouse chain model in ref \cite{key-103}.
Note that a different numerical model might lead to a different numerical
result for $\mu$. For instance, in Fig. 9.21 of Rubinstein and Colby,
the data of a repton model simulation is well approximated by $\mu=1$.
The basic length step of this simulation is equivalent to $N_{p}$
of my work. Since both repton model and my simulations consist of
non-continuous Rouse chains, I expect that $\mu=1$ and $N_{p}$ as
basic step length is more suitable to obtain a quantitative estimate
for the effect of tube length fluctuations.

\bibitem{key-103}M. Doi, \emph{J.Pol.Sci.Pol.Phys.} \textbf{21},
667 (1983).

\bibitem{key-71}M. Müller, J. P. Wittmer, M. E. Cates, \emph{Phys.
Rev E} \textbf{61}, 4078 (2000).

\bibitem{key-72}M. Lang, \emph{Macromolecules} \textbf{46}, 1158,
(2013).

\bibitem{key-100}De Gennes, \emph{Macromolecules} \textbf{19}, 1249
(1986).

\bibitem{key-101}M. Lang, S. Kreitmeier, D. Göritz, \emph{Rubber
Chemistry and Technology} \textbf{80}, 873 (2008).

\bibitem{key-66}Z. Wang, A. E. Likhtman, R. G. Larson, \emph{Macromolecules}
\textbf{45}, 3557 (2012).

\bibitem{key-1}M. Lang, K. Schwenke, J.-U. Sommer, \emph{Macromolecules}
\textbf{45}, 4886 (2012).

\bibitem{key-9}P. Flory, \emph{Proceedings of the Royal Society London
A} \textbf{351}, 351 (1976).

\bibitem{key-41}M. Lang,\emph{ ``Bildung und Struktur in Polymeren
Netzwerken''}, Dissertation, University of Regensburg (2004).

\end{thebibliography}
\end{document}